\documentclass[%
preprint,
amsmath,amssymb,
aps,
]{revtex4-2}

\usepackage{graphicx}
\usepackage{dcolumn}
\usepackage{bm}
\usepackage{hyperref}
\usepackage[dvipsnames]{xcolor}
\usepackage{comment}
\usepackage{multirow}
\usepackage{mathtools}
\usepackage{subcaption}
\usepackage{xcolor}
\usepackage{bigstrut}

\newcommand\fig[1] {Fig.\,{\ref{#1}}}

\def\beq{\begin{equation}}
\def\eeq{\end{equation}}
\def\bsp#1\esp{\begin{split}#1\end{split}}
\def\bal#1\eal{\begin{align}#1\end{align}}

\newcommand\tS   {\theta_\rs}

\newcommand\rL   {\ensuremath{\mathrm{L}}}
\newcommand\rR   {\ensuremath{\mathrm{R}}}

\newcommand\MeV  {\ensuremath{\mathrm{MeV}}}
\newcommand\GeV  {\ensuremath{\mathrm{GeV}}}
\newcommand\rc   {\ensuremath{\mathrm{c}}}
\newcommand\rd   {\ensuremath{\mathrm{d}}}

\newcommand\ri   {\ensuremath{\mathrm{i}}}
\newcommand\rt   {\ensuremath{\mathrm{t}}}
\newcommand\rs   {\ensuremath{\mathrm{s}}}
\newcommand\rF   {\ensuremath{\mathrm{F}}}

\newcommand\Mpl {\ensuremath{{M}_{\rm Pl}}}
\newcommand\mzp {\ensuremath{{M}_Z}}
\newcommand\mwp {\ensuremath{{M}_W}}
\newcommand\mtp {\ensuremath{{M}_{\rm t}}}
\newcommand\mhp {\ensuremath{{M}_h}}
\newcommand\msp {\ensuremath{{M}_s}}

\newcommand\cS {\ensuremath{\cos\theta_\rs}}

\newcommand\sS {\ensuremath{\sin\theta_\rs}}
\newcommand\veff  {\ensuremath{{V}_{\rm eff}}}
\newcommand\ZS {\ensuremath{\mathbf{Z}_\rs}}

\begin{document}

\title{Vacuum stability and scalar masses in the superweak extension of the standard model}
\author{Zolt\'an P\'eli}%
\email{zoltanpeli92@gmail.com}
\affiliation{%
University of Debrecen and \\
ELKH-DE Particle Physics Research Group, 4010 Debrecen, PO Box 105, Hungary
}
\author{Zolt\'an Tr\'ocs\'anyi}
\email{zoltan.trocsanyi@cern.ch}
\affiliation{Institute for Theoretical Physics, ELTE E\"otv\"os Lor\'and University,
P\'azm\'any P\'eter s\'et\'any 1/A, 1117 Budapest, Hungary, also\\
University of Debrecen and \\
ELKH-DE Particle Physics Research Group, 4010 Debrecen, PO Box 105, Hungary}
\date{\today}

\begin{abstract}
We study the allowed parameter space of the scalar sector in the 
superweak extension of the standard model (SM). The allowed region is 
defined by the conditions of (i) stability of the vacuum and 
(ii) perturbativity up to the Planck scale, (iii) the pole mass of 
the Higgs boson falls into its experimentally measured range. We employ
renormalization group equations and quantum corrections at two-loop 
accuracy. We study the dependence on the Yukawa couplings of the 
sterile neutrinos at selected values. We also check the exclusion limit 
set by the precise measurement of the mass of the $W$ boson. Our method
for constraining the parameter space using two-loop predictions
can also be applied to simpler models such as the singlet scalar 
extension of the SM in a straightforward way.
\end{abstract}

\maketitle

\section{Introduction}

Currently particle physics is in a similar situation as physics was about 120 years
ago. Its standard model (SM) can explain successfully most of the low and high energy 
phenomena and provide predictions that are in agreement with measurements at high precision.
Nevertheless, there are also a handful of outstanding observations that cannot be
predicted by the standard model and point towards beyond the standard model (BSM) physics. 
These unexplained facts are
(i) the non-vanishing neutrino masses and mixing matrix elements \cite{Fukuda:1998mi,Ahmad:2001an}, 
(ii) the metastable vacuum of the standard model \cite{Bezrukov:2012sa,Degrassi:2012ry},
(iii) the need for lepto- and/or baryogenesis to explain baryon asymmetry, i.e.~our 
obvious existence, (iv) the existence of dark matter in the Universe 
\cite{Hinshaw:2012aka,Aghanim:2018eyx,Eisenstein:2005su,Sofue:2000jx,Bartelmann:1999yn}, 
and also (v) the existence of dark energy in the Universe \cite{Hinshaw:2012aka}. 
In addition there is general consensus about the occurrence of cosmic inflation 
in the early Universe, which also calls for an explanation. 
There are other observations in particle physics that have almost reached the status
of discoveries. Most prominently the prediction of the standard model for the
anomalous magnetic moment $a_\mu$ of the muon \cite{Aoyama:2020ynm} is smaller 
than the result of the measurement \cite{Muong-2:2006rrc,Muong-2:2021ojo} by 
4.2 standard deviations. In this case however, the status of the theory
is controversial because the evaluation of the hadronic contribution to $a_\mu$ 
requires non-perturbative approach, and the result depends on the method
\cite{Aoyama:2020ynm,Borsanyi:2020mff}. The resolution
of this discrepancy calls for an independent evaluation of this hadronic
vacuum polarization contribution before discovery can be claimed.

Some of the observations (i--v) should find understanding in particle physics models, while 
others may have cosmological origins. Nevertheless, the intimate relation between
particle physics and the early Universe, originating from the universal expansion of
space-time, gives a strong support for searching answers within particle physics by
extending the SM. Such extensions can be put into three categories: 
(a) ultraviolet complete models from theoretical motivations, such as
supersymmetric models; 
(b) effective field theories like the standard model effective field theory (SMEFT); 
(c) simplified models that focus on a subset of open questions.  
This third category includes the dark photon models (gauge extension, see e.g.~Refs.~\cite{Holdom:1985ag,Pospelov:2007mp}), 
the singlet scalar extensions (see e.g.~Refs.~\cite{Schabinger:2005ei,Patt:2006fwx,Falkowski:2015iwa}) and 
the introduction of neutrino mass matrices with some variant of the see-saw mechanism, 
such as in Ref.~\cite{Lindner:2013awa}. 

The UV complete supersymmetric extensions of the SM are very attractive for solving
theoretical issues, but they are becoming less favored by the results of the LHC
experiments \cite{ATLAS:SUSY,CMS:SUSY}. Effective field theories proved to be 
very useful in the past. However, the SMEFT contains 2499 dimension six operators
\cite{Grzadkowski:2010es}, which makes it rather
difficult to study experimentally. The simplified models on the other end
contain only few new parameters, hence are very attractive from the experimental point
of view. However, being simplified models, those cannot give answers to all observations
pointing towards BSM physics simultaneously.

\begin{figure}[t!]
\includegraphics[width=0.85\linewidth]{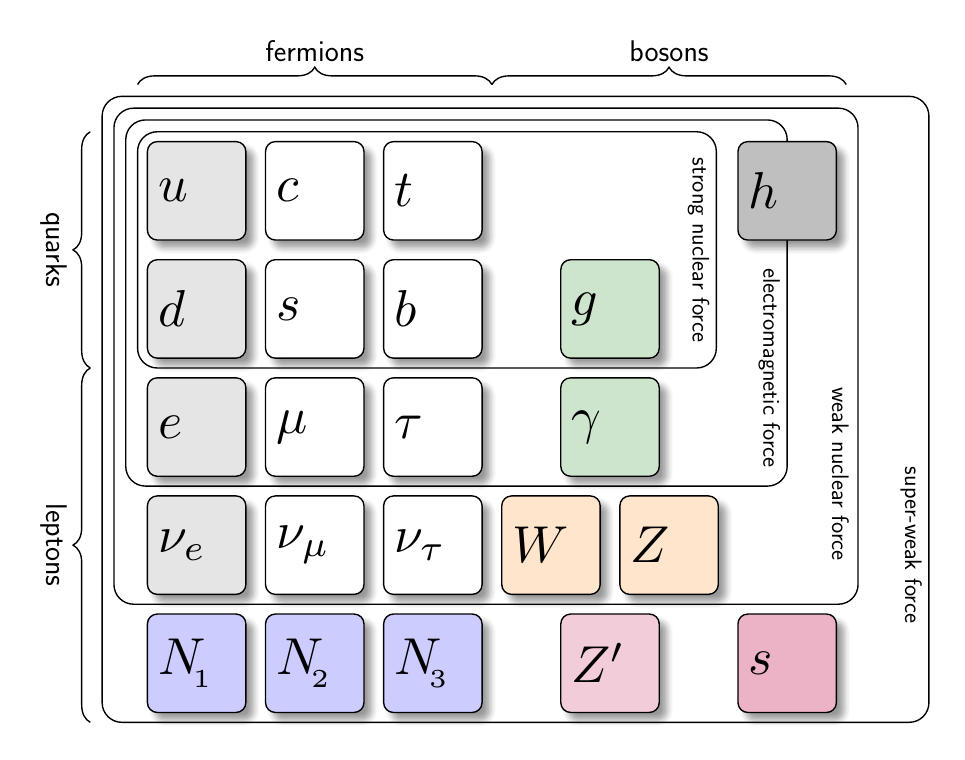}
\caption{\label{fig:forces}The standard model particle sheet with the superweak extension. The forces act on all particles within the respective box
}
\end{figure}

In this paper we study a simple UV complete BSM extension along the principles of the SM
itself: a renormalizable gauge theory that adds one layer of interactions below the 
hierarchic layers of the strong, electromagnetic and weak forces, which is called 
superweak (SW) force \cite{Trocsanyi:2018bkm}, mediated by a new U(1) gauge boson $Z'$, 
see \fig{fig:forces}. In order to explain the origin of neutrino masses, the field 
content is enhanced by three generations of right-handed neutrinos. The new gauge 
symmetry is broken spontaneously by the vacuum expectation value of a new complex 
scalar singlet. According to exploratory studies, the superweak extension of the 
standard model (SWSM) has the potential to explain the origin of 
(i) neutrino masses and mixing matrix elements \cite{Iwamoto:2021wko}, 
(ii) dark matter \cite{Iwamoto:2021fup}, 
(iii) cosmic inflation \cite{Peli:2019vtp}, 
(iv) stabilization of the electroweak vacuum  \cite{Peli:2019vtp} and possibly
(v) leptogenesis (under investigation). 

While these findings are promising, more refined analyses are needed in order to
explore the viability of the model. The main motivation of our work is not to 
prove that the SWSM is the correct description of the fundamental interactions, 
but rather to check if questions (i--v) listed above can be answered within a 
single model with as few new parameters as possible. In this paper we revisit 
the study of the parameter space of the scalar sector of the SWSM as allowed 
by the requirement of the stability of the vacuum. We improve significantly 
on our previous analysis \cite{Peli:2019vtp} in two respects.
Firstly, we use renormalization group equations (RGEs) containing the beta
functions at two-loop order. More importantly, we take into account both 
the radiative corrections up to two-loop accuracy and the measured physical 
values and uncertainties of the parameters of the scalar sector as constraints. 
A similar study has been performed earlier in the simplified model of single 
real scalar extension of the SM in Ref.~\cite{Falkowski:2015iwa}. 
The important difference between the present work and that analysis is that 
we include the effect of the right-handed neutrinos in the running of 
the couplings, which constrains the parameter space further. The inclusion 
of the two-loop effects is also for the first time in the present work.

\section{Superweak model}

The SWSM is a gauged U(1) extension of the standard model with an additional complex
scalar field $\chi$ and three families of sterile neutrinos $\nu_{\rR,i}$.
The model was defined in Ref.~\cite{Trocsanyi:2018bkm} and further details on the
new sectors were presented in Refs.~\cite{Peli:2019vtp,Iwamoto:2021wko}.
Here we recall some details relevant to the present analysis.

The anomaly free charge assignment is shown in Table~\ref{tab:Field-rep}. 
In particular, the $\chi$ field does not couple directly to any fields of the SM.
\begin{table}[th] 
\centering
\caption{\label{tab:Field-rep}Group representations
and charges of the fermions and scalars in the SWSM}
\begin{tabular}{|c|cccc|}\hline\hline
\textbf{field}&     SU(3)$_\rc$ & SU(2)$_\rL$ & U(1)$_Y$ & U(1)$_z$ \\ \hline \hline 
$Q_\rL$       &  \textbf{3} & \textbf{2} & $\frac16$& $\frac16$ \bigstrut\\\hline 
$u_\rR$       & \textbf{3} & \textbf{1} & $\frac23$& $\frac76$ \bigstrut\\\hline 
$d_\rR$       & \textbf{3} & \textbf{1} & $-\frac{1}{3}$& $-\frac{5}{6}$ \bigstrut\\\hline 
$L_\rL$       & \textbf{1} & \textbf{2} & $-\frac12$& $-\frac12$ \bigstrut\\\hline 
$\ell_\rR$    & \textbf{1} & \textbf{1} & $-1$& $-\frac32$ \bigstrut\\\hline 
$N_\rR$       & \textbf{1} & \textbf{1} & $0$& $\frac12$ \bigstrut\\\hline
$\phi$        & \textbf{1} & \textbf{2} & $\frac12$& 1\bigstrut\\\hline 
$\chi$        & \textbf{1} & \textbf{1} & $0$&  $-1$\bigstrut\\\hline 
\hline
\end{tabular}
\end{table}

After spontaneous symmetry breaking (SSB), we parametrize the SM scalar doublet $\phi$ 
and the new scalar field as 
\begin{equation}
\phi = 
\frac{1}{\sqrt{2}}\begin{pmatrix}
-\ri\sqrt{2}\sigma^+\\
v+H+\ri\sigma_\phi
\end{pmatrix},
\quad\text{and}\quad
\chi = \frac{1}{\sqrt{2}}\bigl(w+S+\ri\sigma_\chi\bigr)
\label{eq:parametrization}
\end{equation}
where $v$ and $w$ are the two vacuum expectation values (VEVs), $H$ and $S$ 
are two real, scalar fields and $\sigma^+$, $\sigma_{\phi/\chi}$ are charged 
and neutral Goldstone bosons. In terms of these fields the scalar potential 
in the SWSM is given by
\begin{equation}
\label{eq:V_phichi}
    V(\phi,\chi) = V_0 - \mu_\phi^2 |\phi|^2 - \mu_\chi^2 |\chi|^2
    +\lambda_\phi |\phi|^4 + \lambda_\chi |\chi|^4
    +\lambda |\phi|^2|\chi|^2
    \,.
\end{equation}
The constant $V_0$ is irrelevant in our considerations, so we set it to zero in the rest of the paper. 
Substituting the parametrization \eqref{eq:parametrization} into \eqref{eq:V_phichi}, we obtain
the tree-level (effective) potential
\begin{equation}\label{eq:V(h,s)}
V(H,S) = 
- \frac12\Big(\mu_\phi^2 H^2 + \mu_\chi^2 S^2\Big)
+ \frac14\Big(\lambda_\phi H^4 + \lambda_\chi S^4 + \lambda~H^2 S^2\Big)
\end{equation}
of the real scalar fields.  The VEVs are determined by the tadpole equations:
\begin{equation}
\label{eq:tadpole}
\bsp
  \frac{\partial V}{\partial H}\biggl|_{H=v,S=w} = 0 &=
  v\biggl( -\mu_\phi^2 + \frac{1}{2}\lambda w^2 + \lambda_\phi v^2\biggr),
  \\
\frac{\partial V}{\partial S}\biggl|_{H=v,S=w} = 0 &=
  w\biggl( -\mu_\chi^2 + \frac{1}{2}\lambda v^2 + \lambda_\chi w^2\biggr)
  \,.
 \esp
\end{equation}
The mass matrix of the scalar fields is given by the Hessian:
\begin{equation}
\label{eq:hessian}
    \textbf{M}_\rs^2 = 
    \begin{pmatrix}
    \frac{\partial^2 V}{\partial H^2} & \frac{\partial^2 V}{\partial H\, \partial S} \\
    \frac{\partial^2 V}{\partial S\, \partial H} & \frac{\partial^2 V}{\partial S^2}
    \end{pmatrix}_{H=v,S=w}
    =\begin{pmatrix}
    2\lambda_\phi v^2 & \lambda v w \\
    \lambda v w       & 2\lambda_\chi w^2
    \end{pmatrix}\,,
\end{equation}
which can be diagonalized by a rotation
matrix
\begin{equation}
    \ZS = \begin{pmatrix}
    \cS & \sS \\
    -\sS & \cS
    \end{pmatrix},
\end{equation}
so that $\ZS^T \textbf{M}_\rs^2 \ZS = \text{diag}(\mhp^2,\msp^2)$. The
parameters $\mhp$ and $\msp$ are the masses of the propagating states $h$ and $s$
\footnote{We shall denote the pole mass of a particle $p$ as $M_p$.}.
The positivity condition for the masses implies the condition 
\beq
(4 \lambda_\chi \lambda_\phi - \lambda^2) v^2 w^2 > 0
\label{eq:positivity}
\eeq
among the scalar couplings and VEVs. Explicitly, the angle of rotation and the scalar masses 
$\mhp$ and $\msp$ can be expressed through the VEVs and couplings at tree level as
\begin{eqnarray}
\label{eq:thetas_tree}
  \tan(2\tS) &=& \frac{\lambda v w}{\lambda_\chi w^2 - \lambda_\phi v^2},\\
  \label{eq:mh_tree}
  \mhp^2 &=& 
  \lambda_\phi v^2 + \lambda_\chi w^2 - \frac{\lambda_\chi w^2 - \lambda_\phi v^2}{\cos(2\tS)}\,\,,
  \\
  \label{eq:ms_tree}
  \msp^2 &=& 
  \lambda_\phi v^2 + \lambda_\chi w^2 + \frac{\lambda_\chi w^2 - \lambda_\phi v^2}{\cos(2\tS)}\,. 
\end{eqnarray}
In the absence of mixing ($\lambda=0$, $\tS=0$) we have
$\mhp = \sqrt{2\lambda_\phi v^2}$, $\msp = \sqrt{2\lambda_\chi w^2}$. 
As the scalar fields are coupled to the $W^\pm$ bosons with the 
interaction vertices 
\begin{equation}
    \Gamma_{hWW}^{\mu\nu} = \frac{\ri}{2}\biggl(g_\rL^2 v \cS\biggr)g^{\mu\nu}\,,
    \quad\text{and}\quad
    \Gamma_{sWW}^{\mu\nu} = \frac{\ri}{2}\biggl(g_\rL^2 v \sS\biggr)g^{\mu\nu}\,,
\end{equation}
only the BEH field is coupled to the $W$ bosons and to the other SM fields
in the limit of vanishing mixing between the scalars. Hence, we naturally
identify the VEV $v$ as that related to the Fermi coupling and also 
the parameter $\mhp$ with the mass of the Higgs boson measured 
at the LHC \cite{ParticleDataGroup:2020ssz} by introducing the notation
\begin{equation} \label{eq:mhcond}
    m_h = 125.10\,\GeV,\quad 
    \Delta m_h = 0.14\,\GeV    \quad\text{and}\quad
    v = \Big(\sqrt{2}G_\rF\Big)^{-1/2} = 246.22\,\GeV\,,
\end{equation}
and requiring $\mhp \in [m_h-\Delta m_h,m_h+\Delta m_h]$. In accordance
with this assumption, we restrict $\tS$ to fall in the range 
$(-\pi/4,\pi/4)$.

The VEV $w$ can be expressed through these known parameters and the scalar couplings using
Eqs.~\eqref{eq:thetas_tree} and \eqref{eq:mh_tree},
\begin{equation}\label{eq:wev_tree}
w = \mhp \sqrt{\frac{\mhp^2 - 2 \lambda_\phi v^2}
{2\lambda_\chi\bigl(\mhp^2 - 2 \lambda_\phi v^2 \bigr)+ \lambda^2 v^2}}
\,.
\end{equation}
Thus, the formal conditions for the non-vanishing $w$, required at the electroweak 
scale are either 
\beq
\mhp^2 > 2 \lambda_\phi v^2
\,,\quad\text{with}\quad
4 \lambda_\chi \lambda_\phi > \lambda^2
\eeq
(the second condition deriving from the positivity constraint in \eqref{eq:positivity} for positive $v^2 w^2$), or
\begin{equation}\label{eq:wvev_cond}
4 \lambda_\chi \biggl(\lambda_\phi - \frac{1}{2}\frac{\mhp^2}{v^2} \biggr) > \lambda^2\,,\quad\text{if}\quad 2 \lambda_\phi v^2 > \mhp^2> 0\,.
\end{equation}
As we have fixed $v$ and $\mhp$ experimentally, the input value of 
$\lambda_\phi$ decides which of these two cases are to be considered.

Eqs.~\eqref{eq:V(h,s)}--\eqref{eq:hessian} are valid at tree level. The effect of the quantum 
corrections can be summarized by substituting the potential $V$ with the effective potential 
$\veff$, whose formal loop expansion is.
\beq
\veff = \sum_{i=0}^\infty \veff^{(i)}
\label{eq:Veff-expansion}
\eeq
where $\veff^{(0)} = V$ and $\veff^{(i)}$ represents the $i$-loop correction.

\section{Vacuum stability in the SWSM at one-loop accuracy}

The potential (\ref{eq:V(h,s)}) is stable if it is bounded from below.
Due to its continuity in the field variables, it is sufficient to study 
the positivity of (\ref{eq:V(h,s)}) for large values of $h$ and $s$, 
which translates to the following conditions on the quartic scalar
couplings:
\begin{equation}
\bsp
\label{eq:stab_conditions}
  \lambda_\phi,\lambda_\chi &> 0\,,\\
   4\lambda_\phi \lambda_\chi - \lambda^2 &> 0  \quad\mbox{for}\quad \lambda<0\,. 
\esp
\end{equation}
Taking into account the radiative corrections leads to 
(i) dependence on the renormalization scale $\mu$ for all renormalized couplings and 
(ii) the corrections $\veff^{(i)}$. While it is straightforward to require that the 
conditions (\ref{eq:stab_conditions}) be satisfied for the running couplings at 
any sensible value of $\mu$, we cannot write the stability conditions for the one-loop
effective potential in a closed form such as in Eq.~(\ref{eq:stab_conditions}) valid 
at tree level. Instead, we take an alternative path by requiring the existence 
of a non-vanishing $w(\mtp)$ indirectly, extracting it from the known pole mass
of the Higgs boson, rather than computing it explicitly form the effective potential
\eqref{eq:Veff-expansion} with radiative corrections taken into account.
Our procedure can be described in terms of analytic expressions at the one-loop 
accuracy as follows.

\begin{figure}[t!]
\includegraphics[width=0.475\linewidth]{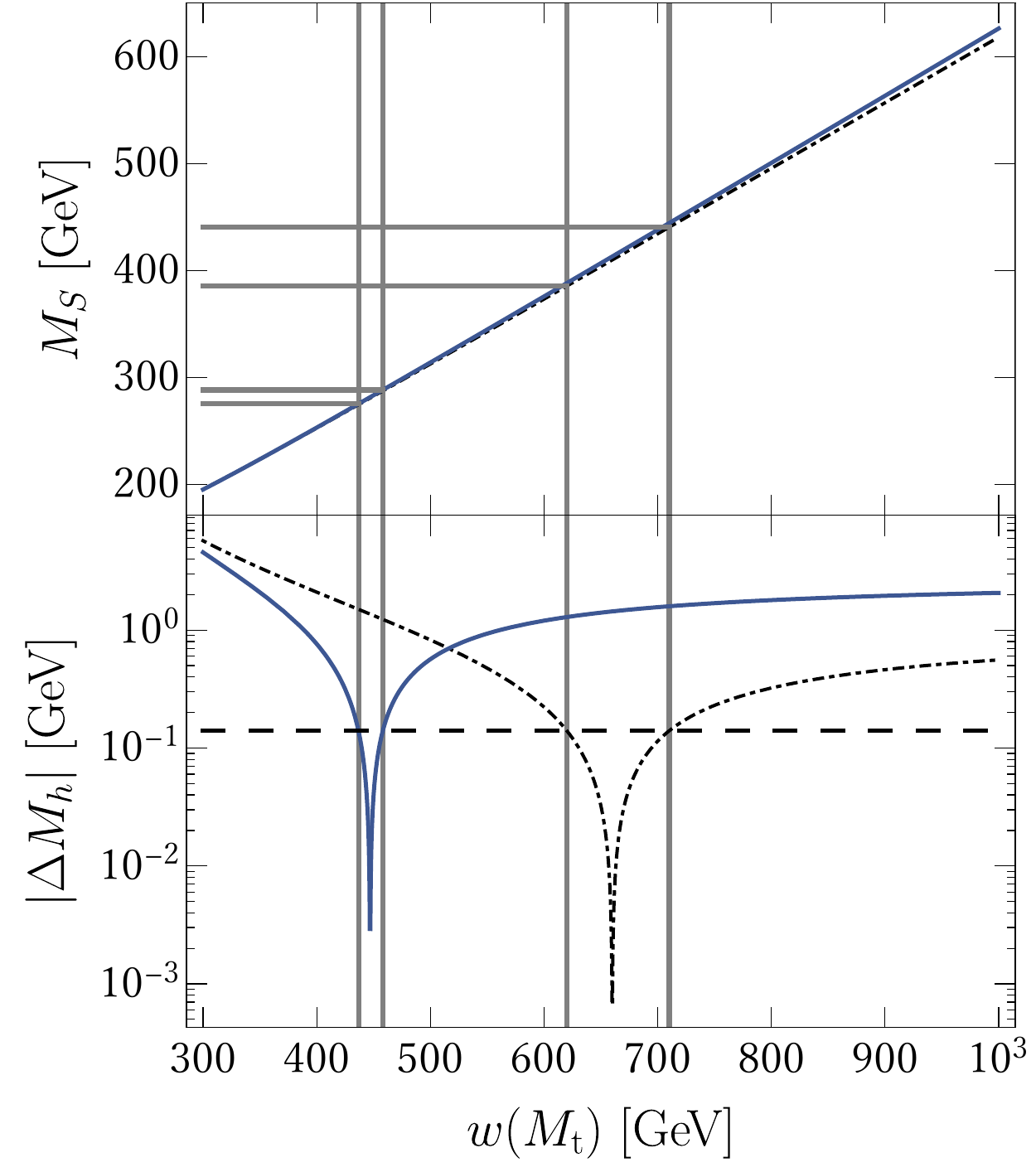}
\includegraphics[width=0.475\linewidth]{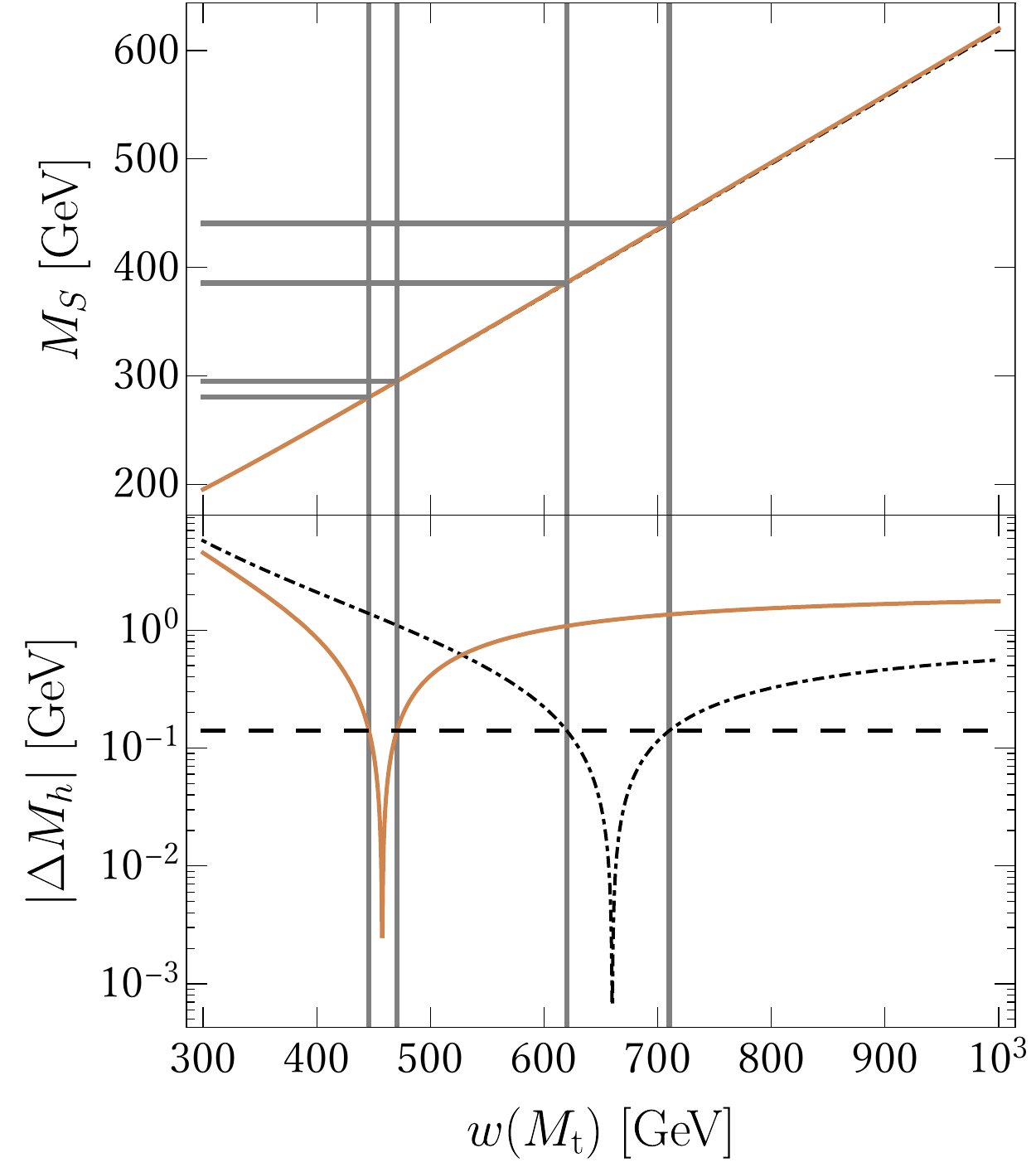}
\caption{\label{fig:wdependence} 
Dependence of the absolute difference $\mhp$ minus the observed Higgs boson mass 
on $w(\mtp)$ (bottom) and the dependence of $\msp$ on $w(\mtp)$ (top)
with input values $\lambda_\phi(\mtp)=0.15$, $\lambda_\chi(\mtp)=0.2$ and 
$\lambda(\mtp)=0.1$.
Left: $y_x(\mtp) = 0$, right: $y_x(\mtp) = 0.8$. The dashed horizontal line corresponds 
to the uncertainty $\Delta m_h$. The black dash-dotted curves are computed at tree level 
(Eqs.~\eqref{eq:mh_tree} and \eqref{eq:ms_tree}), while the solid colored ones at one loop 
(Eqs.~\eqref{eq:mh_1loop} and \eqref{eq:ms_1loop}). 
}
\end{figure}

We investigate the vacuum stability in the range $\mu\in (\mtp, \Mpl)$, i.e.~from the
pole mass $\mtp$ of the t quark up to the Planck mass $\Mpl$ where quantum
gravitational effects become important. The scale dependence of a given coupling $g$ 
is described by the autonomous system coupled differential equations of the form
\begin{equation}
    \frac{\partial g}{\partial t} = \beta_g\,,
\end{equation}
called RGEs, where $\partial/\partial t = \mu \,\partial/\partial \mu$.
We assume that the model remains perturbatively valid for the complete 
range by requiring 
\begin{equation}\label{eq:pt_conditions}
    |g(\mu)| < 4 \pi\,,\quad \mu\in (\mtp, \Mpl)
\end{equation}
for any coupling $g$ in the theory, which we check in the stability analysis.
Consequently, we can employ perturbation theory to compute the $\beta_g$
functions. We integrate the complete set of RGEs of the SWSM, while requiring the
stability and perturbativity conditions \eqref{eq:stab_conditions} and
\eqref{eq:pt_conditions}. We also assume the existence of $w$ at the scale $\mu=\mtp$, which implies the existence of a second
massive neutral gauge boson and a second massive scalar particle as predictions of 
the model. To check this condition, we compute the loop corrected scalar mixing
angle and scalar pole masses:
\begin{eqnarray}
\label{eq:thetas_1loop}
  &&\tan\bigl(2\tS(p^2)\bigr) = \frac{\lambda(\mu) v(\mu) w(\mu) + \Pi_{HS}(p^2)}{\lambda_\chi(\mu) w(\mu)^2 - \lambda_\phi(\mu) v(\mu)^2+ \Pi_{-}(p^2)},
  \\
  \label{eq:mh_1loop}
  &&\mhp^2 =
  \lambda_\phi(\mu) v(\mu)^2 + \lambda_\chi(\mu) w(\mu)^2 +\Pi_{+}(\mhp^2) 
  - \frac{\lambda_\chi(\mu) w(\mu)^2 - \lambda_\phi(\mu) v(\mu)^2 +\Pi_{-}(\mhp^2)}{\cos\bigl(2\tS(\mhp^2)\bigr)}\,,
  \\
  \label{eq:ms_1loop}
  &&\msp^2 = \lambda_\phi(\mu) v(\mu)^2 + \lambda_\chi(\mu) w(\mu)^2 +\Pi_{+}(\msp^2)
  + \frac{\lambda_\chi(\mu) w(\mu)^2 - \lambda_\phi(\mu) v(\mu)^2 +\Pi_{-}(\msp ^2)}{\cos\bigl(2\tS( \msp^2)\bigr)}\,,
\end{eqnarray}
using the shorthand notation
\begin{equation}
 \Pi_{\pm}(p^2) = \frac{1}{2}\biggl(\tilde{\Pi}_{SS}(p^2) \pm\tilde{\Pi}_{HH}(p^2)\biggr)\,,
\end{equation}
where $\tilde{\Pi}_{\varphi\varphi}(p^2) = \Pi_{\varphi\varphi}(p^2)-T_\varphi/\langle\varphi\rangle$, with 
$\Pi_{\varphi_I \varphi_J}(p^2)$ being the sum of all one particle irreducible 
(1PI) two-point
functions with external legs $\varphi_I$ and $\varphi_J$, while $T_\varphi$ is the
sum of all 1PI one-point functions with external leg $\varphi$ ($\varphi$, 
$\varphi_I = H$ or $S$). In other words,
Eqs.~\eqref{eq:thetas_1loop}--\eqref{eq:ms_1loop} are valid at any order 
in perturbation theory. We collect these one- and two-point functions 
computed at one-loop accuracy in App.~\ref{app:oneloop}. As shown explicitly, 
each coupling and VEV in Eqs.~\eqref{eq:thetas_1loop}--\eqref{eq:ms_1loop} 
depends on the renormalization scale $\mu$, but the pole masses $\mhp^2$ 
and $\msp^2$ do not up to the effect of neglected higher order corrections. 
An important check of our calculations is the independence of the scalar pole 
masses $\mhp$ and $\msp$ of the renormalization scale $\mu$
\begin{equation}\label{eq:pole_scaling}
\mu \frac{\partial \mhp}{\partial \mu } =   \mu \frac{\partial \msp}{\partial \mu } = 0.
\end{equation}
As mentioned, we identify the pole mass $\mhp$, computed in perturbation theory in
\eqref{eq:mh_1loop} as the observed Higgs boson mass $m_h\pm \Delta m_h$, which 
constrains the possible values of $w(\mtp)$ severely for a given set of input 
couplings at $\mu = \mtp$. The lower panels in Fig.~\ref{fig:wdependence}  
show the dependence of $|\Delta M_h|$, with $\Delta M_h = M_h-m_h$, 
on $w(\mtp)$. We see that it falls below the experimental uncertainty 
$\Delta m_h$, represented by the dashed lines, in a fairly narrow range of 
$w(\mtp)$. To find the range of values of the allowed $w^{(i)}(\mtp)$, 
with superscript referring to the accuracy in the perturbative order,
we solve the two equations 
\begin{equation}
    \mhp(w^{(1)})\biggr|_{\mu=\mtp} = m_h\pm\Delta m_h
\end{equation}
for $w^{(1)}(\mtp)$ numerically. We consider the two solutions physical
if those are positive, shown by the vertical lines. Then we use the accepted values 
$w^{(1)}(\mtp)$, falling into the ranges between the vertical line,
to compute the possible values of $\msp$ using Eq.~\eqref{eq:ms_1loop}. This
procedure is shown by the plots on the top of Fig.~\ref{fig:wdependence} 
for a specific set of input couplings.

The complete set of running couplings can be grouped into three sets. The 
(i) SM couplings $g_Y,~g_\rL,~g_\rs,~y_\rt$, the
(ii) SW gauge coupling $g_z$ and 
(iii) the scalar quartic couplings $\lambda_\phi,\lambda_\chi,\lambda$ 
together with the sterile neutrino Yukawa coupling $y_x$. We assume one light
sterile neutrino -- a candidate for dark matter \cite{Iwamoto:2021fup} -- and two 
heavy ones with equal masses for simplicity, $y_x = y_{x,5} = y_{x,6}$. 
We neglect the effect of the SW gauge coupling from our analysis because 
its maximally allowed value is very small, $g_z \lesssim 10^{-4}$, if the model 
is to explain the origin of dark matter \cite{Iwamoto:2021fup} and also should 
obey the direct observational limit of the NA61 experiment \cite{NA64:2019imj}.
Explicitly, in group (iii) we have the following autonomous system of RGEs at one loop:
\begin{equation}\label{eq:1loop_RGE}
\bsp
  \frac{\partial \lambda_\phi}{\partial t} &= \beta_{\lambda_\phi,\text{SM}}^{(1)}+ \frac{\lambda^2}{(4\pi)^2}\,,\quad
  \frac{\partial \lambda_\chi}{\partial t} =
  \frac{1}{(4\pi)^2}\biggl(20\lambda_\chi^2+2\lambda^2 - 2 y_x^4 +4\lambda_\chi y_x^2 \biggr),
  \\
  \frac{\partial \lambda}{\partial t} &=    
  \frac{\lambda}{(4\pi)^2}\biggl(-\frac{3}{2}g_Y^2 -
  \frac{9}{2} g_\rL^2 +12\lambda_\phi + 8 \lambda_\chi +4\lambda +6 y_t^2 + 2y_x^2\biggr),
\esp
\end{equation}
for the scalar couplings, with $\beta_{\lambda_\phi,\text{SM}}^{(1)}$ being the 
one-loop beta function of the SM quartic scalar coupling, and
\begin{equation}\label{eq:1loop_RGEy}
  \frac{\partial y_x}{\partial t} = 
  \frac{2 y_x^3}{(4\pi)^2}\,,
  \qquad
  \frac{\partial w}{\partial t} = -\frac{ w}{(4\pi)^2} \frac{y_x^2}2
\end{equation}
for the Yukawa coupling and new VEV. The one-loop beta functions show, that 
a sufficiently large Higgs portal coupling $\lambda$ is able to drive $\lambda_\phi$ 
and $\lambda_\chi$ to positive values, while the sterile neutrino Yukawa couplings
drive $\lambda_\chi$ towards negative values. The last equation, the RGE for $w$ does
not affect the vacuum stability analysis. We present it as it is used in checking
the conditions in Eq.~\eqref{eq:pole_scaling}.

There are three SM precision parameters measured precisely, $G_\rF,~\mzp$ and
$\alpha_{\text{em}}^{\overline{\text{MS}}}(\mzp)$, which can be turned into input 
values for the couplings in group (i) together with the less precisely known $\mtp$ 
and $\alpha_\rs^{\overline{\text{MS}}}(\mzp)$.
The self energies $\Pi_{WW}(p^2)$ and $\Pi_{ZZ}(p^2)$ also receive
contributions $\Pi^{\text{SW}}_{VV}(p^2)$ due to the SW extension, 
given in Eq.~\eqref{eq:wz_selfenergy}, which shift the input values 
of the VEV $v$ and the electroweak gauge couplings.
Hence, we use the following inputs in group (i)
\begin{equation}
    \bsp
     g_Y(\mtp) &= 0.3586 +  \delta g_Y(\mtp),
     \\
     g_\rL(\mtp) &= 0.6477 +  \delta g_\rL(\mtp),
     \\
     v(\mtp) &= 247.55~\GeV + \delta v(\mtp),
    \esp
\end{equation}
with $g_\rs(\mtp)=1.167$ and $y_\rt(\mtp)=0.940$. The SW corrections 
$\delta g_Y,\delta g_\rL$ and $\delta v$ are defined in
App.~\ref{app:electroweak-correction}. The SM value of the gauge and scale 
dependent VEV $v$ in the Feynman gauge is $v_{\text{SM}}(\mtp) = 247.55~\GeV$. 
The SW corrections to the electroweak input parameters $\delta g_Y$,
$\delta g_\rL$ and $\delta v$ are small and do not modify noticably 
our final results even at two loops. We take the value of $y_\rt(\mtp)$ 
from the fit formula (25) of Ref.~\cite{Degrassi:2012ry} as the
largest possible value. This choice is the most conservative one concerning the
vacuum stability because the main culprit causing the metastable SM vacuum is the 
large value of the t quark Yukawa coupling $y_t(\mtp)$. The last set (iii) of the 
input couplings are unconstrained and we scan their values at $\mu=\mtp$ in order 
to obtain the parameter space in $\{\lambda_\phi,\lambda_\chi,\lambda,y_x\}_{\mu=\mtp}$ 
where the stability \eqref{eq:stab_conditions}, perturbativity \eqref{eq:pt_conditions} 
conditions in the range $\mu\in (\mtp, \Mpl)$, together with existence of 
the $w$ vacuum at $\mu= \mtp$ are fulfilled.

\begin{figure}[t!]
{
\includegraphics[width=0.47\linewidth]{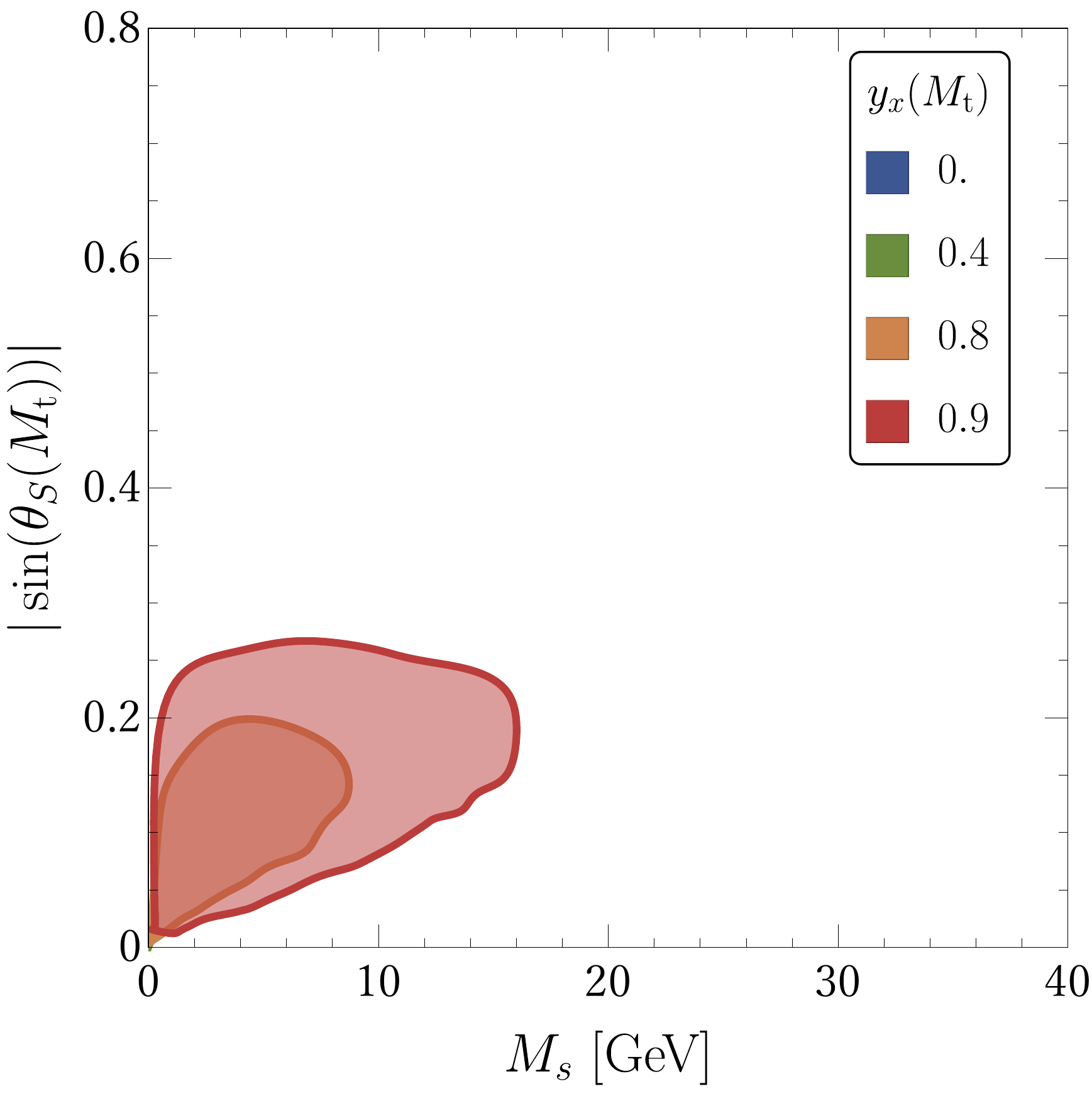}
\includegraphics[width=0.48\linewidth]{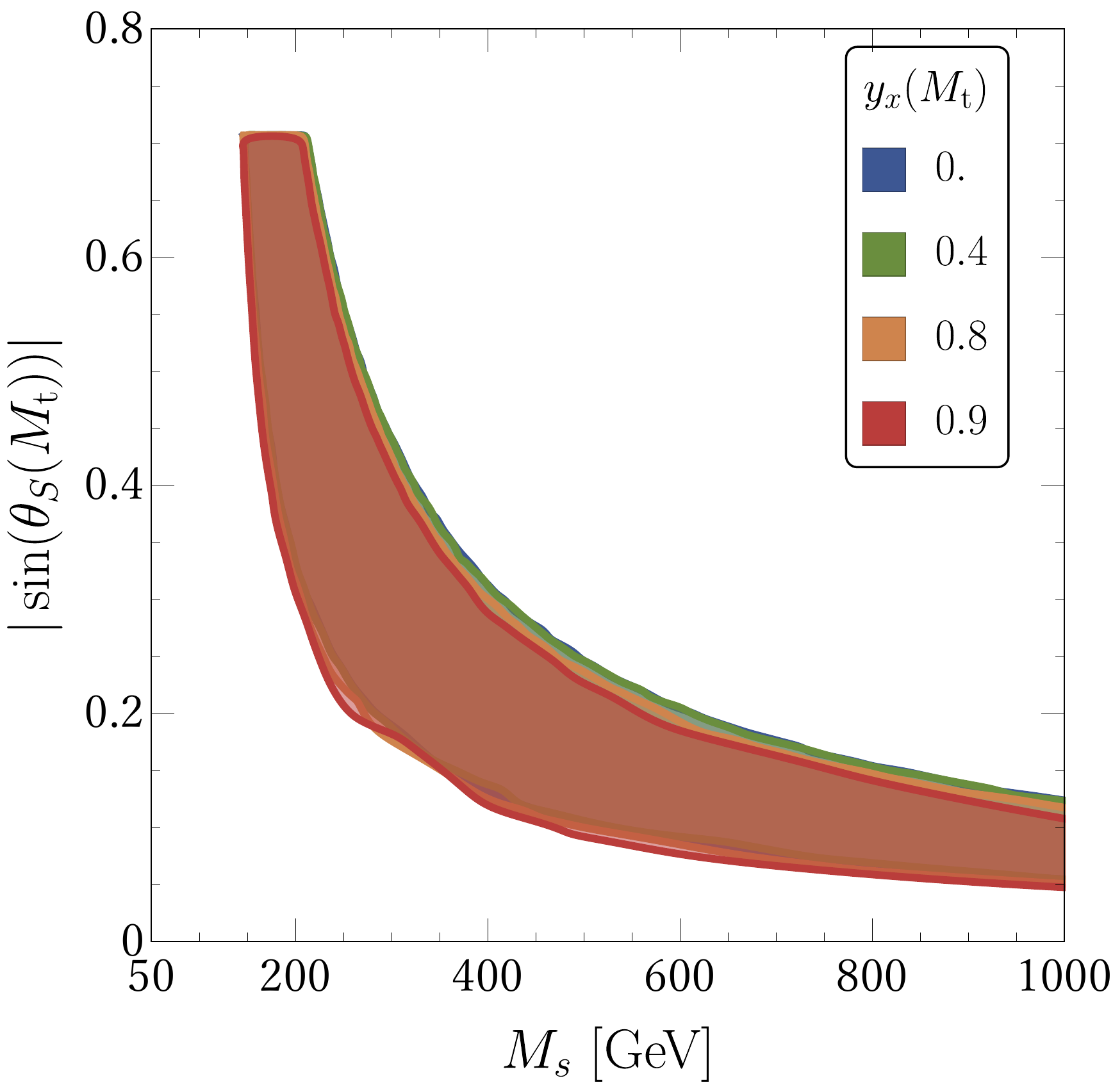}
}
\caption{\label{fig:1loop_masses} 
Allowed parameter space $V_\lambda(y_x)$ in the $\msp-|\sin(\Theta_S)|$ plane 
at representative values of $y_x$ at one loop accuracy.
The different colored areas correspond to different values of $y_x(\mtp)$
as shown in the legends. Left: $\msp < \mhp$, right: $\msp > \mhp$. 
}
\end{figure}

We have scanned the volume $V_\lambda(y_x)= \{ \lambda_\phi,\lambda_\chi,\lambda \}_{\mu=\mtp}$ 
spanned by the input couplings at fixed values of $y_x(\mtp)$ to find the parameter
space allowed by our conditions. There are two quantitatively different regions. 
In the first one (a) $\msp < \mhp$, i.e.~the new scalar is lighter than the 
Higgs boson, whereas in the second one (b) $\msp > \mhp$. We shall present the 
result of such scans in the next section where we the computations will be performed 
at two-loop accuracy. Having found the allowed region of the input parameters, 
we can compute the scalar mixing angle and mass of the new scalar using
Eqs.~\eqref{eq:thetas_1loop} and \eqref{eq:ms_1loop}, to obtain the allowed parameter 
space in the $\msp-|\sin(\tS(\mtp))|$ plane, shown in Fig.~\ref{fig:1loop_masses} 
at selected values of the neutrino Yukawa coupling. 
In case (a), the parameter space is empty for $y_x(\mtp) = 0$, but it grows 
non-linearly with increasing $y_x(\mtp)$. For instance, the parameter space 
for $y_x(\mtp) = 0.4$ is not empty, but still invisible at the resolution of
Fig.~\ref{fig:1loop_masses}, as in that case one has $\msp < 300\,\MeV$ and 
$|\sin(\tS)| < 0.04$. For $y_x(\mtp) \gtrsim 1$ the stability condition 
$\lambda_\chi >0$ is not satisfied at any scale below $\Mpl$. It turns out 
that in case (b) the value of the VEV $w^{(1)}(\mtp)$ and hence that of 
$\msp$ can be larger than shown in the plot when the scalar mixing coupling 
tends to zero. As that also means vanishing mixing coupling $\lambda$, 
it represents the phenomenologically rather irrelevant case of very weakly 
coupled dark sector. 

\section{Vacuum stability in the SWSM at two-loop accuracy}\label{sec:2loops}

In order to check the robustness of the perturbative analysis of the parameter 
space where the vacuum is stable, we have repeated the procedure described in the 
previous section at two-loop accuracy. Given a set of input couplings
$\{\lambda_\phi(\mtp),\lambda_\chi(\mtp),\lambda(\mtp),y_x(\mtp)\}$, we first 
computed $w^{(1)}(\mtp)$ at $\mu=\mtp$, using our analytic formulae as
described in the previous section. We solved the two-loop $\beta$-functions to check the
conditions of stability and perturbativity only if we found $w^{(1)}(\mtp) > 0$. 
If all the stability and perturbativity conditions were fulfilled for the input 
values  $\{\lambda_\phi(\mtp),\lambda_\chi(\mtp),\lambda(\mtp),y_x(\mtp)\}$, we used
\texttt{SPheno} \cite{Porod:2003um,Porod:2011nf} to compute the scalar pole masses 
at two-loops, $\mhp^{(2)}$ and $\msp^{(2)}$, using  $w$ as a free input 
parameter. Starting from the initial value $w = w^{(1)}(\mtp)$, 
we searched for the $w$ at which 
\beq
\mhp^{(2)}(w) = m_h\,,
\eeq
which we call $w=w^{(2)}$. This procedure of starting with using only such points 
in the parameter space where the condition $w^{(1)}(\mtp) > 0$ is satisfied 
saves significant CPU time as the numerical solution of the two-loop 
$\beta$-functions and especially the computations in \texttt{SPheno} are very time consuming
\footnote{There is a price to pay for this speed-up, namely we discard 
small portions of the parameter space, where $w^{(1)}(\mtp)$ is not positive 
but $w^{(2)}(\mtp)$ is so.}.
\begin{figure}[t!]
\includegraphics[width=0.5\linewidth]{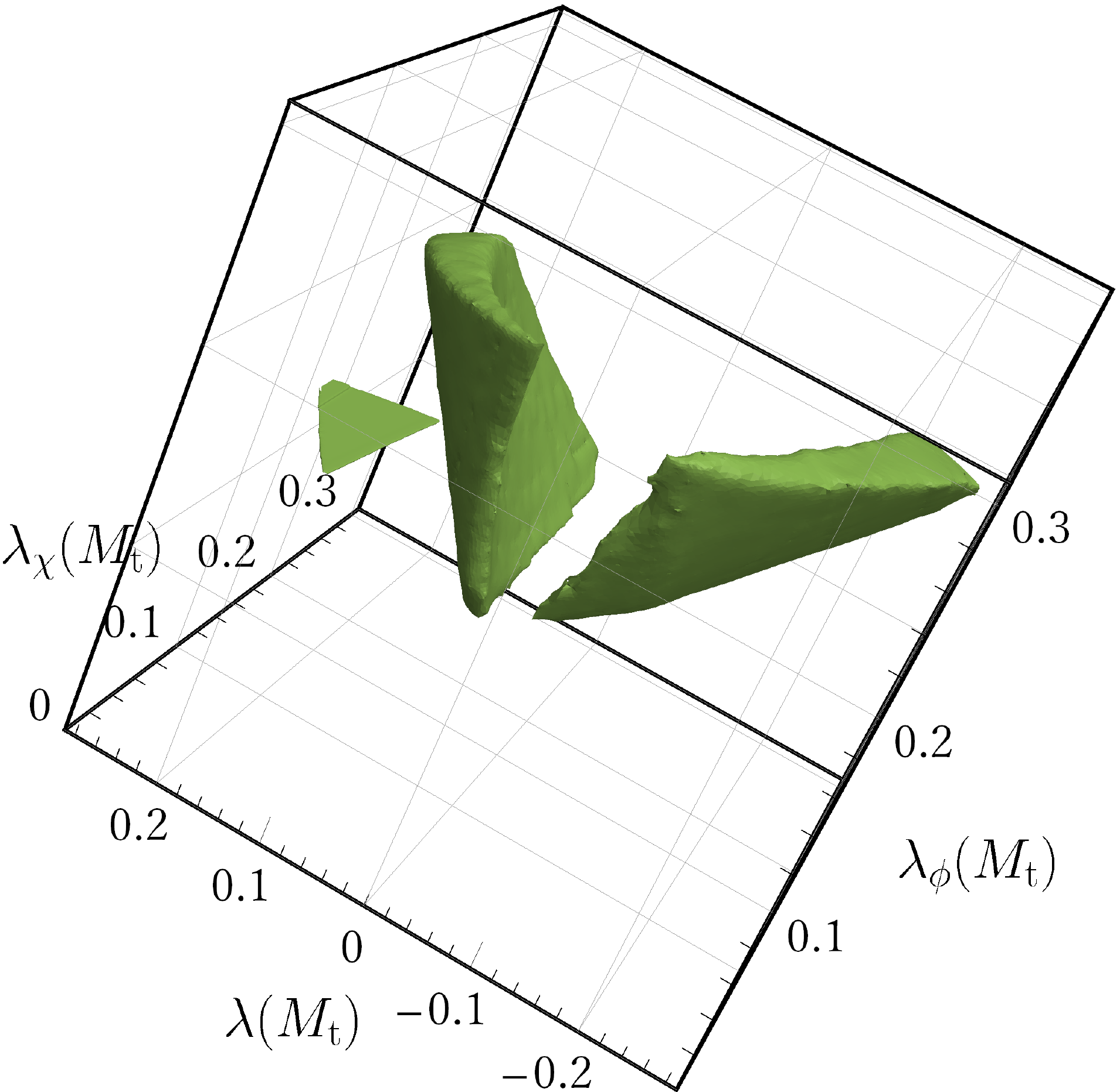}
\caption{\label{fig:3d param} Three dimensional parameter space at $y_x(\mtp) = 0.4$.
}
\end{figure}
\begin{figure}[t!]
{
\includegraphics[width=0.47\linewidth]{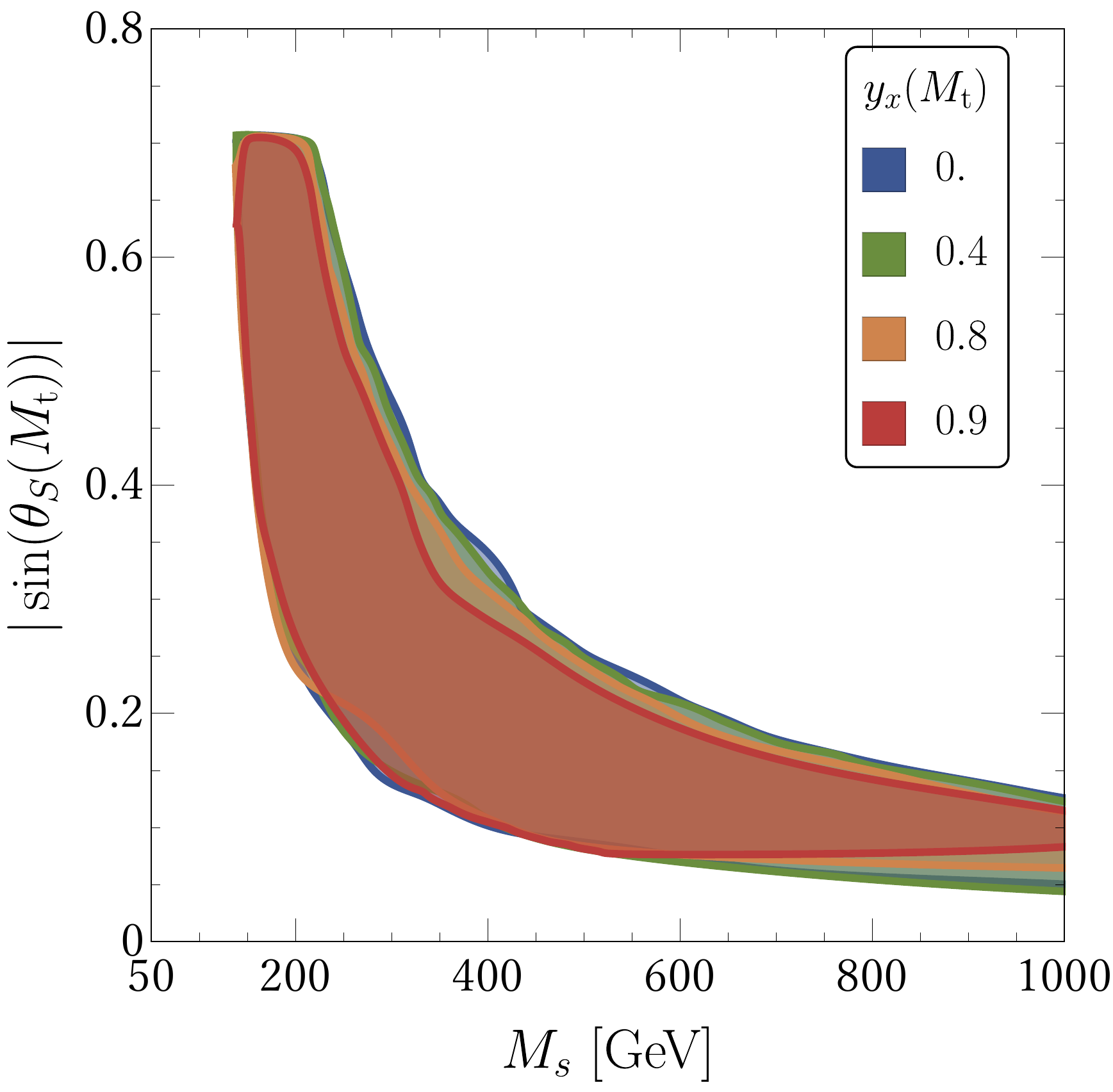}
\includegraphics[width=0.485\linewidth]{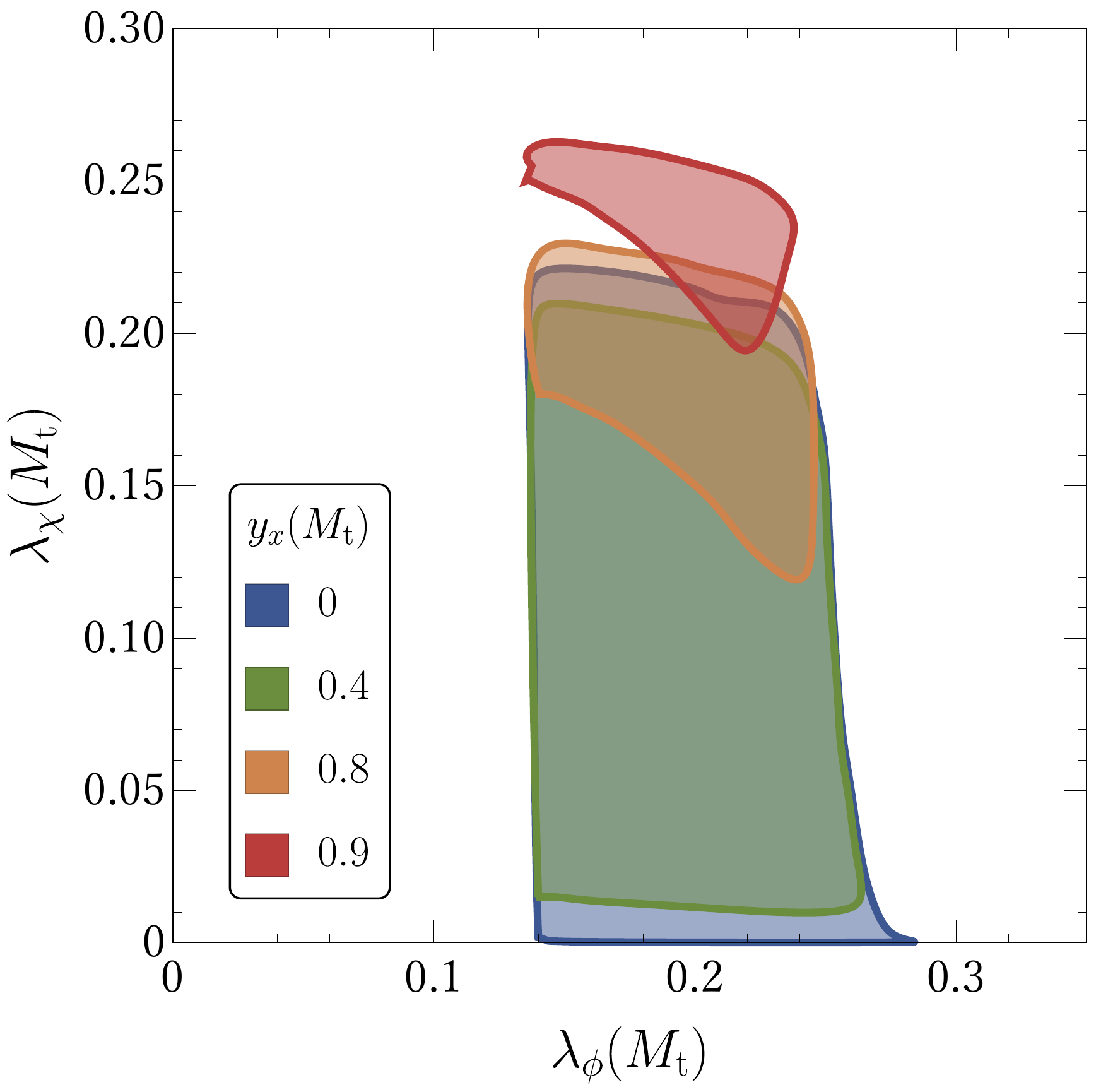}
}
\\
{
\includegraphics[width=0.48\linewidth]{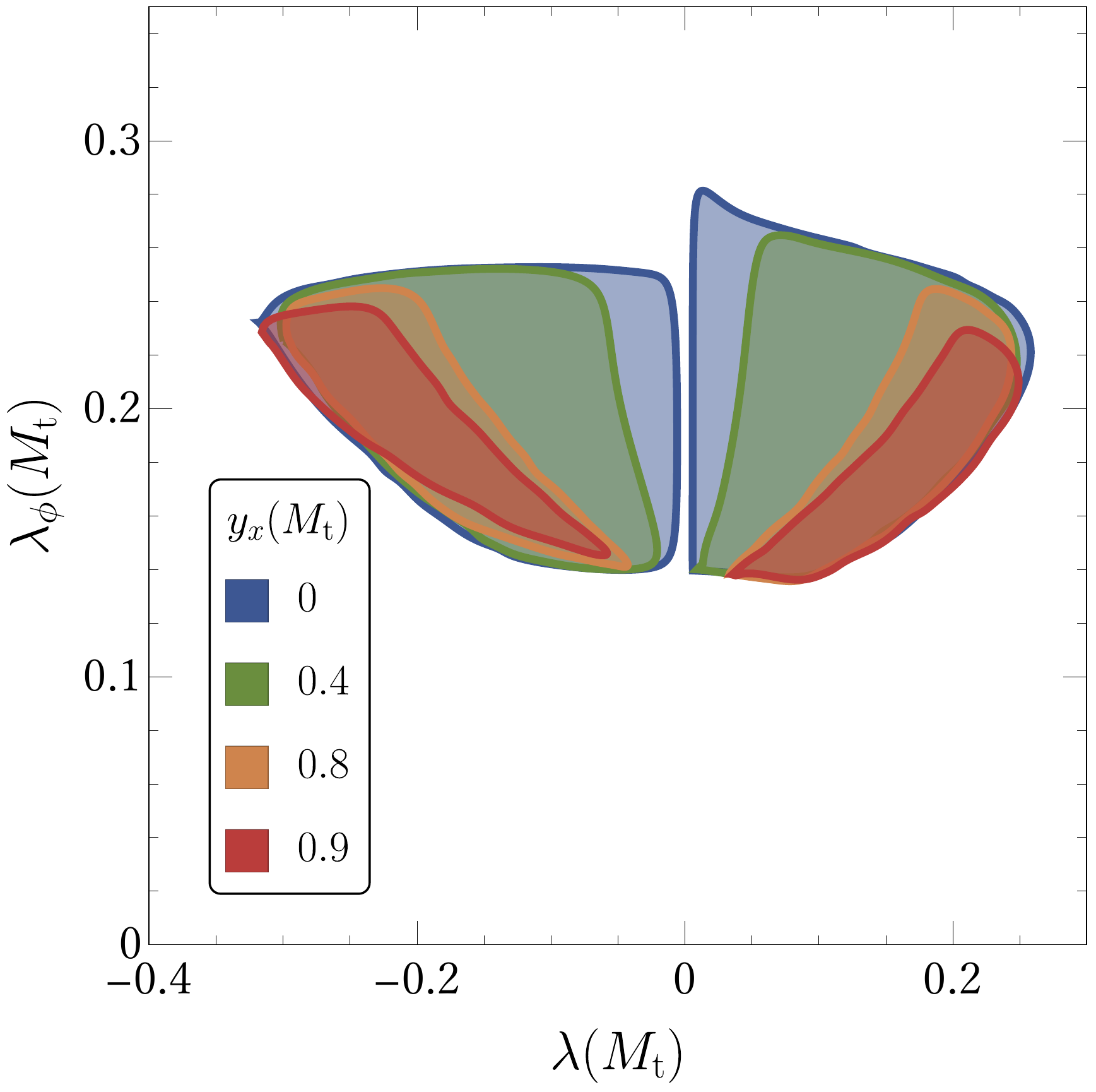}
\includegraphics[width=0.48\linewidth]{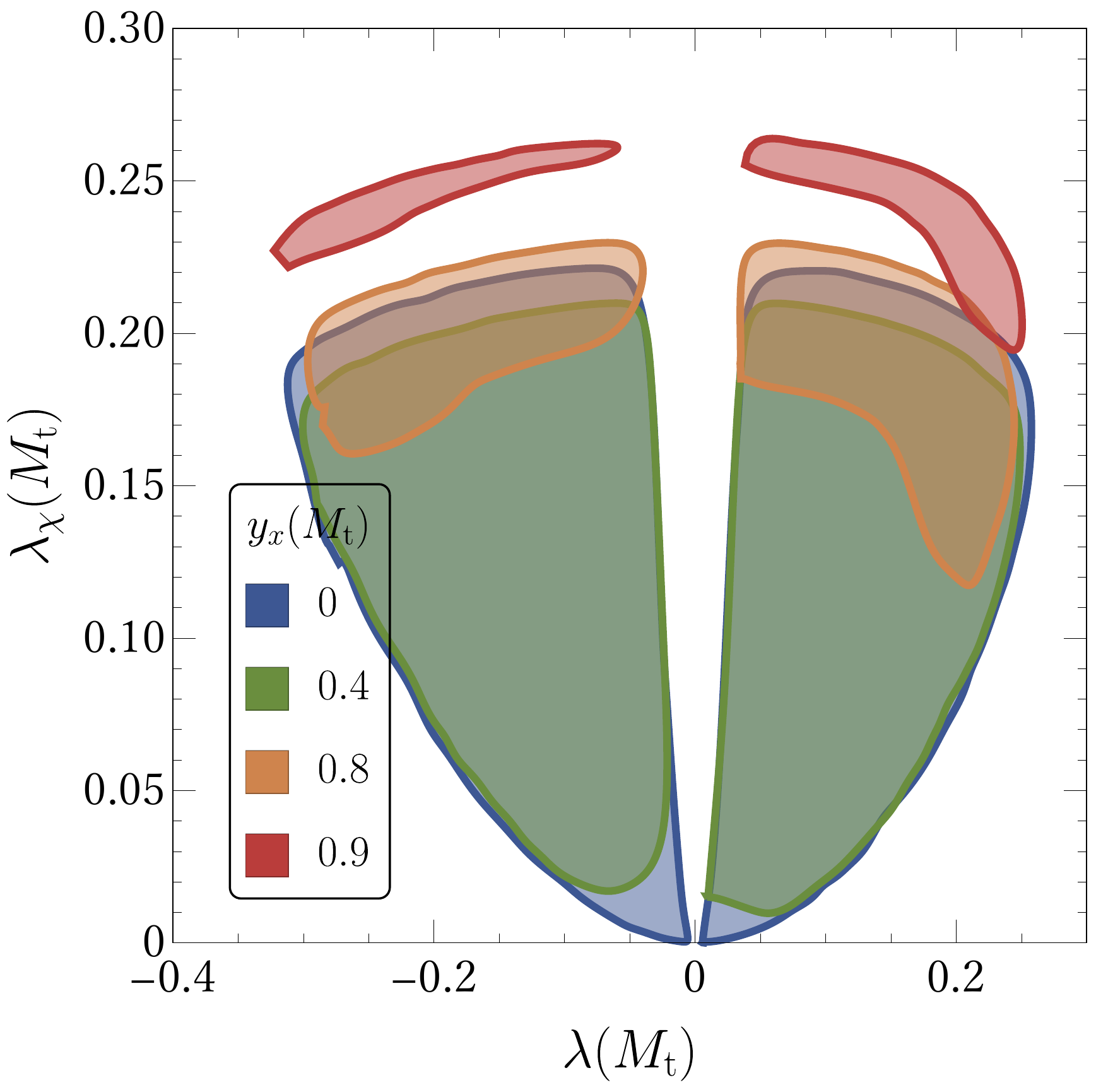}
}
\caption{\label{fig:lambdaspace_point2} 
Planar projections of the allowed  parameter space, where $\msp > \mhp$
and the conditions \eqref{eq:stab_conditions}, \eqref{eq:pt_conditions}, 
$w^{(2)}(\mtp) > 0$ are fulfilled at two-loop accuracy. Top left: allowed regions in the 
$\msp-|\sin\tS|$ plane, other plots show the two-dimensional projections of the 
three-dimensional allowed regions in $V_\lambda(y_x)$. The different colored 
regions correspond to different  values of $y_x$ as shown in the legends.
}
\end{figure}

\begin{figure}[t!]
\includegraphics[width=0.48\linewidth]{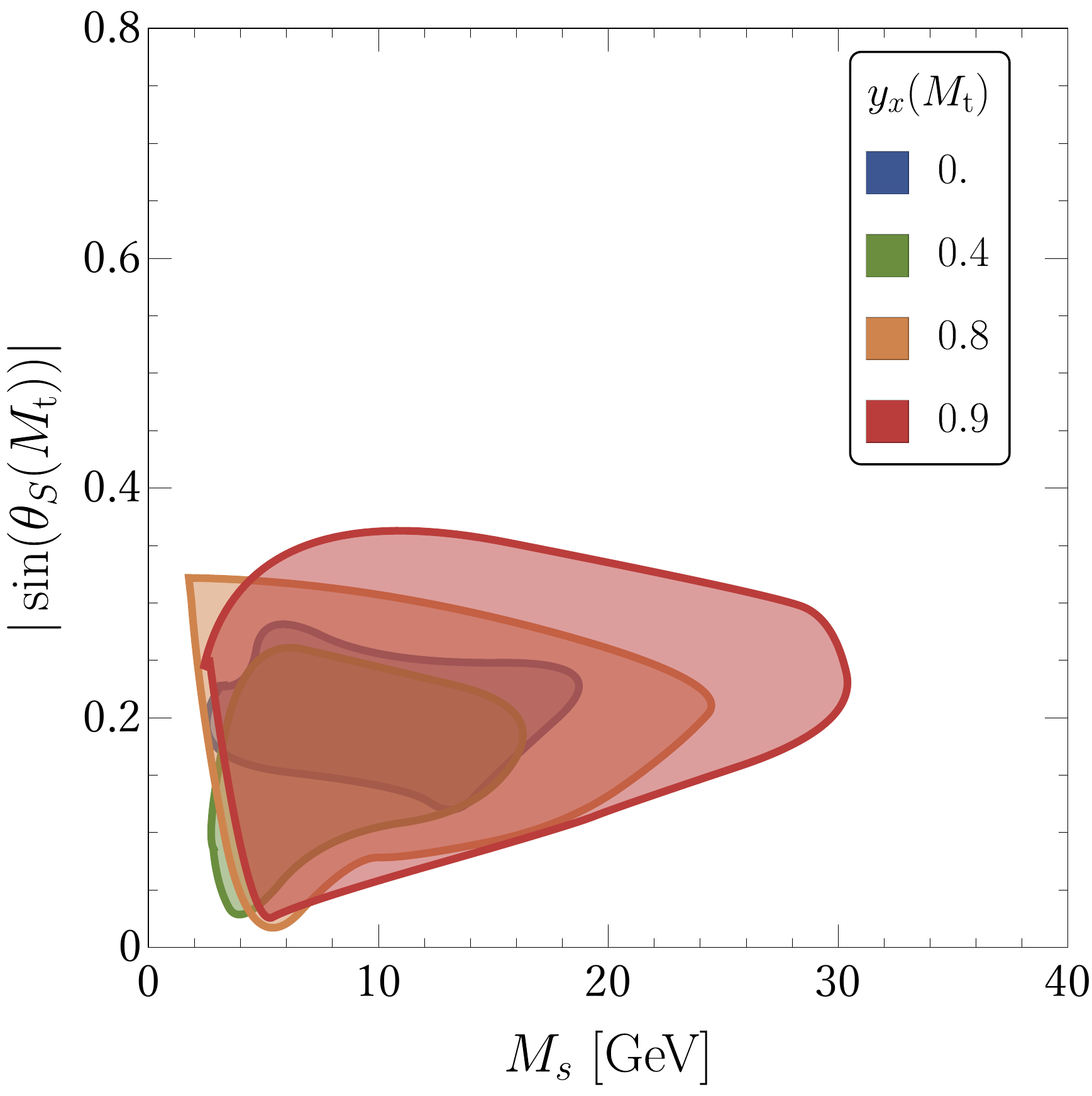}
\includegraphics[width=0.475\linewidth]{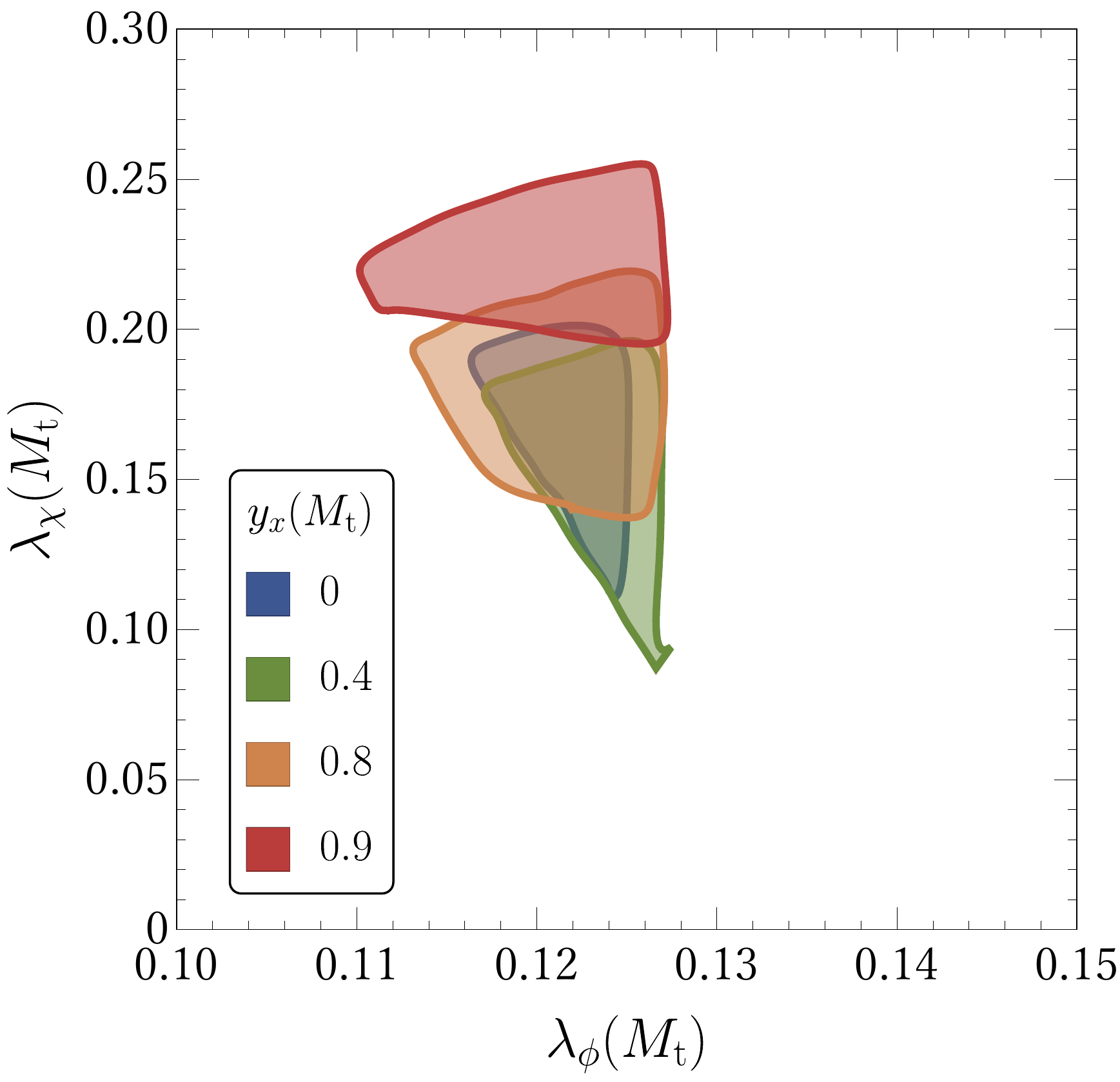}
\\
\includegraphics[width=0.48\linewidth]{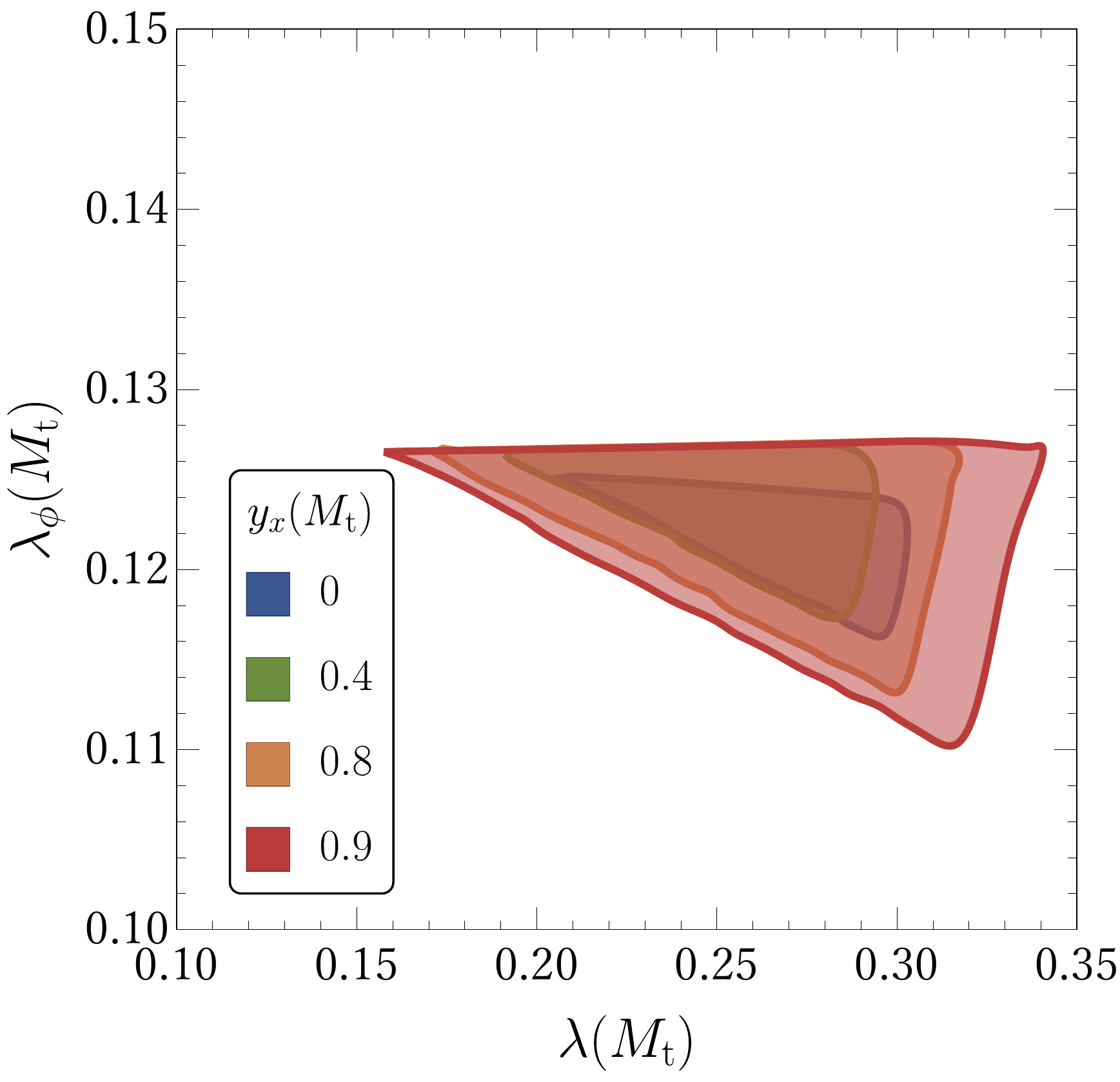}
\includegraphics[width=0.49\linewidth]{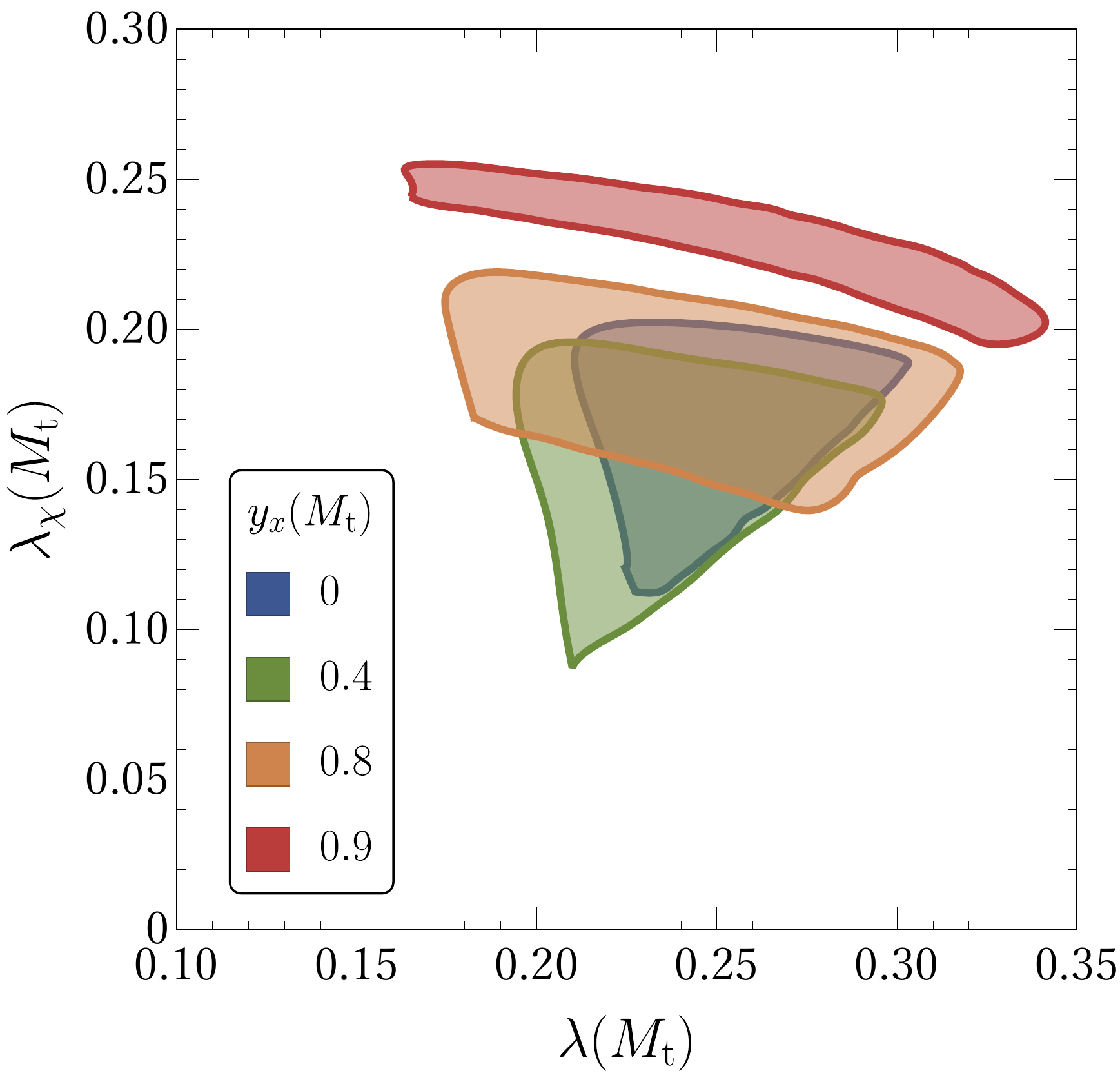}
\caption{\label{fig:lambdaspace_point1} 
Same as Fig.~\ref{fig:lambdaspace_point2} for $\msp > \mhp$.
}
\end{figure}

The parameter space is shown by a perspective view at $y_x(\mtp) = 0.4$ in 
Fig.~\ref{fig:3d param}, and its projections to the two-dimensional sub-spaces 
at selected values of $y_x(\mtp)$ in Figs.~\ref{fig:lambdaspace_point2} and 
\ref{fig:lambdaspace_point1}. We have also computed the regions $V_\lambda(y_x)$ 
using the tree-level relation \eqref{eq:wev_tree} instead of the one-loop one in
Eq.~\eqref{eq:mh_1loop} as done typically in the case of scalar singlet extensions
(see e.g.~\cite{SM-eft-stability}). We found that the allowed region on the
$\msp-|\sin(\tS)|$ plane for case (a) $\msp < \mhp$ is sensitive to such a change 
in an interesting way: for vanishing Yukawa coupling the allowed region found using 
Eq.~\eqref{eq:wev_tree} disappears at one loop (see the discussion of Fig.~\ref{fig:1loop_masses} in the previous section), 
but reappears at two loops, which had not been found in pervious analyses.
If $y_x(\mtp)$ is increased from zero, we find 
non-empty parameter space at any of the first three orders in perturbation theory. 
The minimum value of $\msp$ in region (a) depends on $y_x(\mtp)$, but it is always 
larger than about 1\,GeV in the two-loop analysis.

One can make two important remarks about the parameter space in case 
(b) $\msp > \mhp$ presented in Fig.~\ref{fig:lambdaspace_point1}.
On the one hand, the parameter space shrinks as $y_x(\mtp)$ increases 
and it disappears completely for $y_x(\mtp) \gtrsim 1$.
On the other hand, we have $|\lambda(\mtp)|>0$, 
because for $\lambda(\mtp)=0$ the scalar mixing vanishes, and then 
$\lambda_\phi$ coincides with $\lambda_{\rm SM}$, which does not satisfy
the vacuum stability conditions, while preserving the pole mass of the Higgs boson.
Also, the volume $V_\lambda(y_x)$ increases slightly with increasing order 
in perturbation theory.

\begin{figure}[t!]
\includegraphics[width=0.475\linewidth]{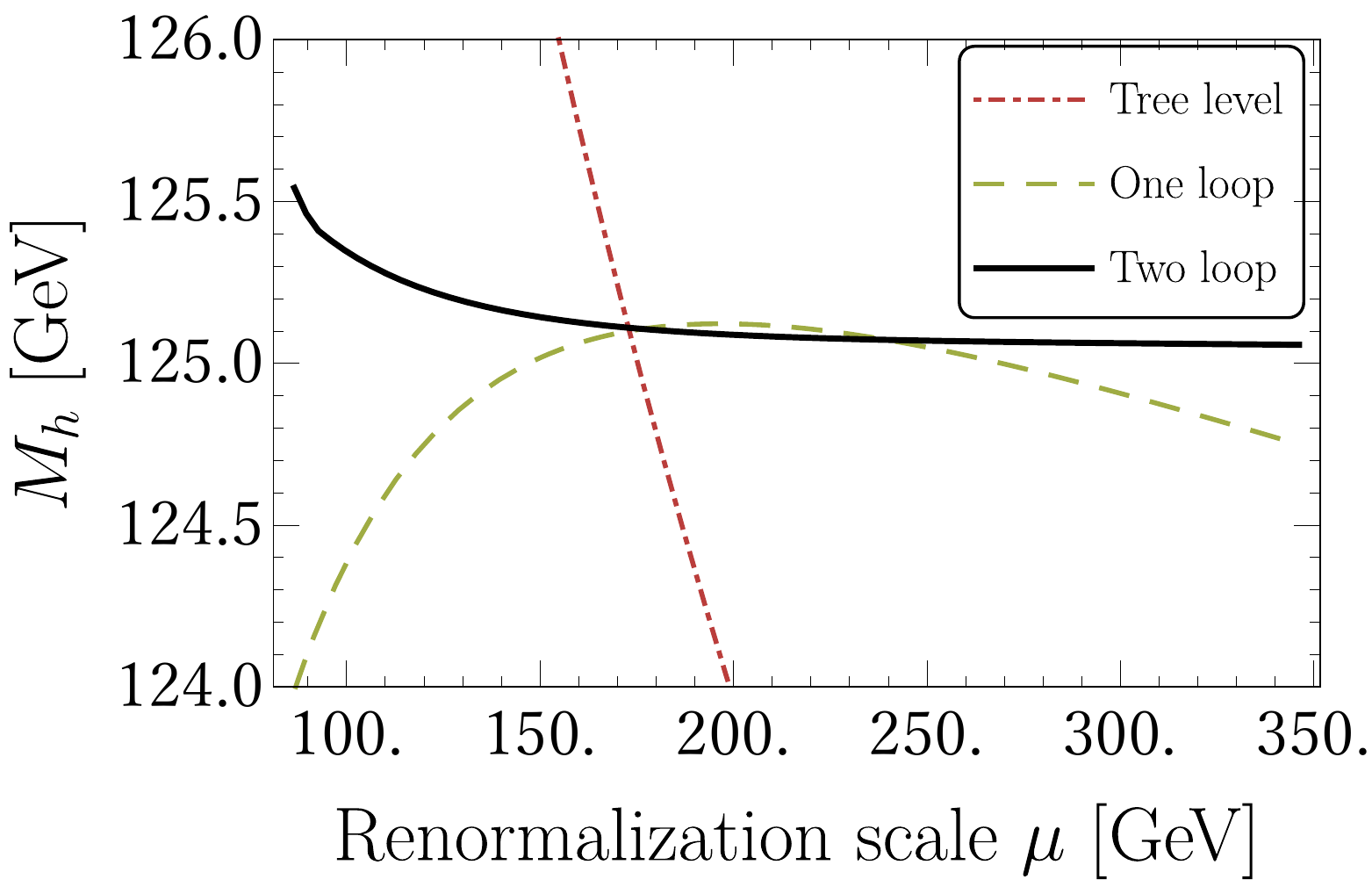}
\includegraphics[width=0.475\linewidth]{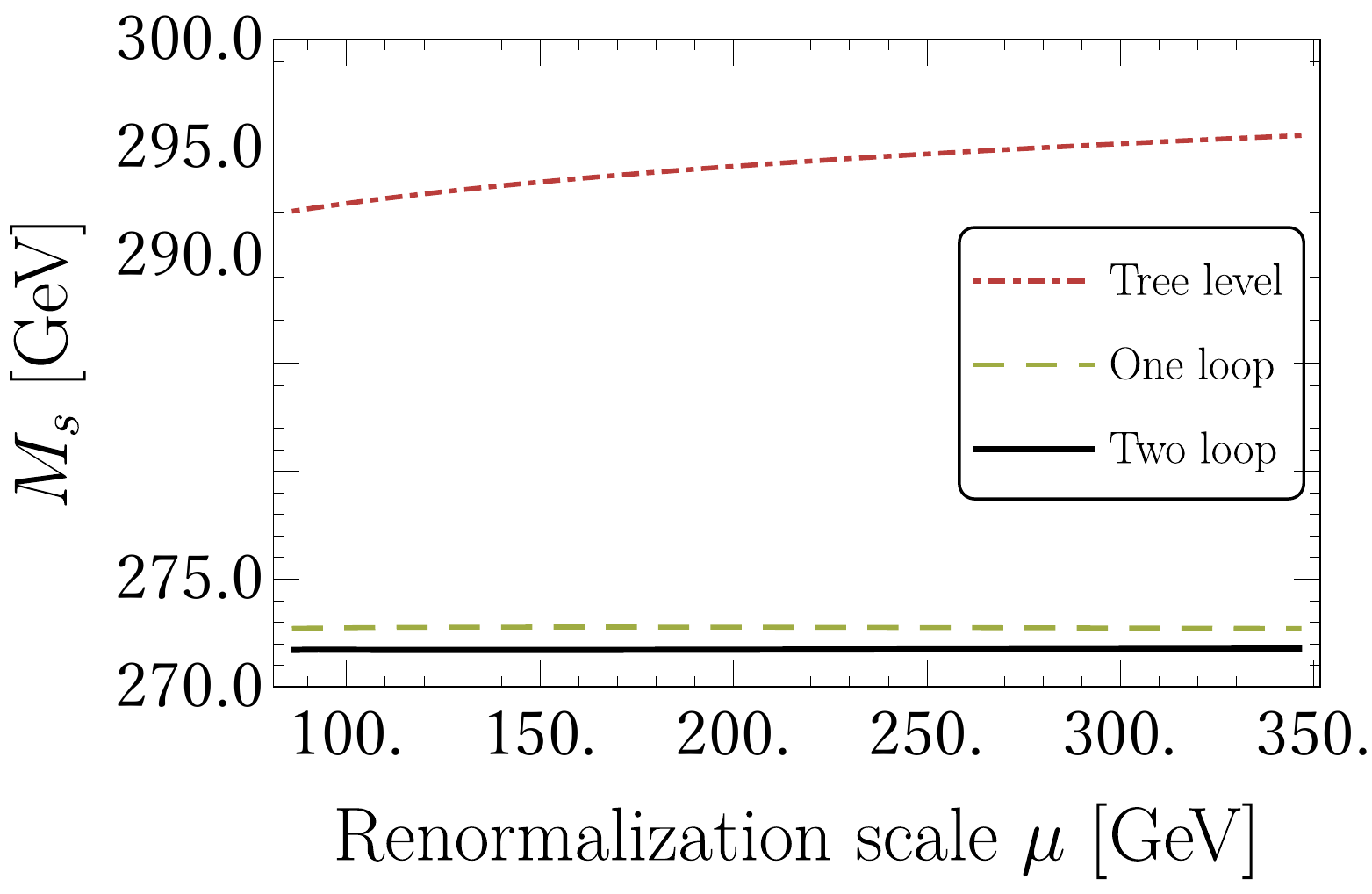}
\caption{\label{fig:scale_check} Dependencees of the scalar boson masses $\mhp^{(i)}$ and 
$\msp$ on the renormalization scale $\mu$ in the range $(0.5\mtp,2\mtp)$ at different 
orders of the perturbation theory at $y_x(\mtp)=0.4$, 
$\lambda_\phi(\mtp)=0.241$, $\lambda_\chi(\mtp) = 0.096$, $\lambda(\mtp)=0.217$.
}
\end{figure}

We have checked Eq.~(\ref{eq:pole_scaling}) numerically for both cases
(a) and (b) at randomly selected input values
$\{\lambda_\phi,\lambda_\chi,\lambda,y_x\}_{\mu=\mtp}$ in the range 
$\mu\in\bigl(0.5\mtp,2\mtp \bigr)$ and compared the scale 
dependences of the tree level masses \eqref{eq:mh_tree} and 
\eqref{eq:ms_tree} to the scale dependences of the one-loop 
accurate pole masses (\ref{eq:mh_1loop}) and (\ref{eq:ms_1loop}). 
As shown in Fig.~\ref{fig:scale_check}, we have found  that the scale 
dependences of the tree level masses are reduced significantly at one-loop 
and even more at two-loop accuracy. The sizeable difference between the scalar 
pole masses $\msp$ at the first two orders of perturbation theory (and to much less 
extent between the next two orders) are not caused by radiative corrections. 
Rather than loop corrections to the masses, the jumps originate from the shifts in
$w(\mtp)$ required to reproduce Higgs boson pole mass at different orders of 
perturbation theory, as can be seen in Fig.~\ref{fig:wdependence}.

The theoretical prediction for the $W$-boson mass uses precision electroweak 
observables (except $\mwp$ itself) and it is sensitive to new physics 
\cite{Robens:2015gla}. Hence, it is often used as a benchmark compared 
to the experimentally observed value \cite{ParticleDataGroup:2020ssz}
\begin{equation}
    M_W^{\text{exp.}} = 80.379 \pm 0.012\,\GeV\,.
\end{equation}
The current, most precise theoretical estimates are 
$M_W^{\text{theo.}} = 80.359 \pm 0.011\,\GeV$ \cite{Baak:2012kk}, 
$80.362 \pm 0.007\,\GeV$ \cite{Ciuchini:2013pca} and 
$80.357 \pm 0.009 \pm 0.003\,\GeV$ \cite{degrassi2014}. 
We take 80.360\,GeV as SM prediction and the combined uncertainty from 
the experimental and theoretical values to be 
$\Delta\mwp \simeq 17\,\MeV$. We set twice this value as an upper limit
to find the allowed range for the new physics contribution to 
$M_W^{\text{theo.}}$, which means that the SW contribution to the 
mass of the $W$ boson  is excluded outside the range $(19\pm 34)$\,MeV, 
i.e~outside $[-15,53]$\,MeV.

\begin{figure}[t!]
\includegraphics[width=0.47\linewidth]{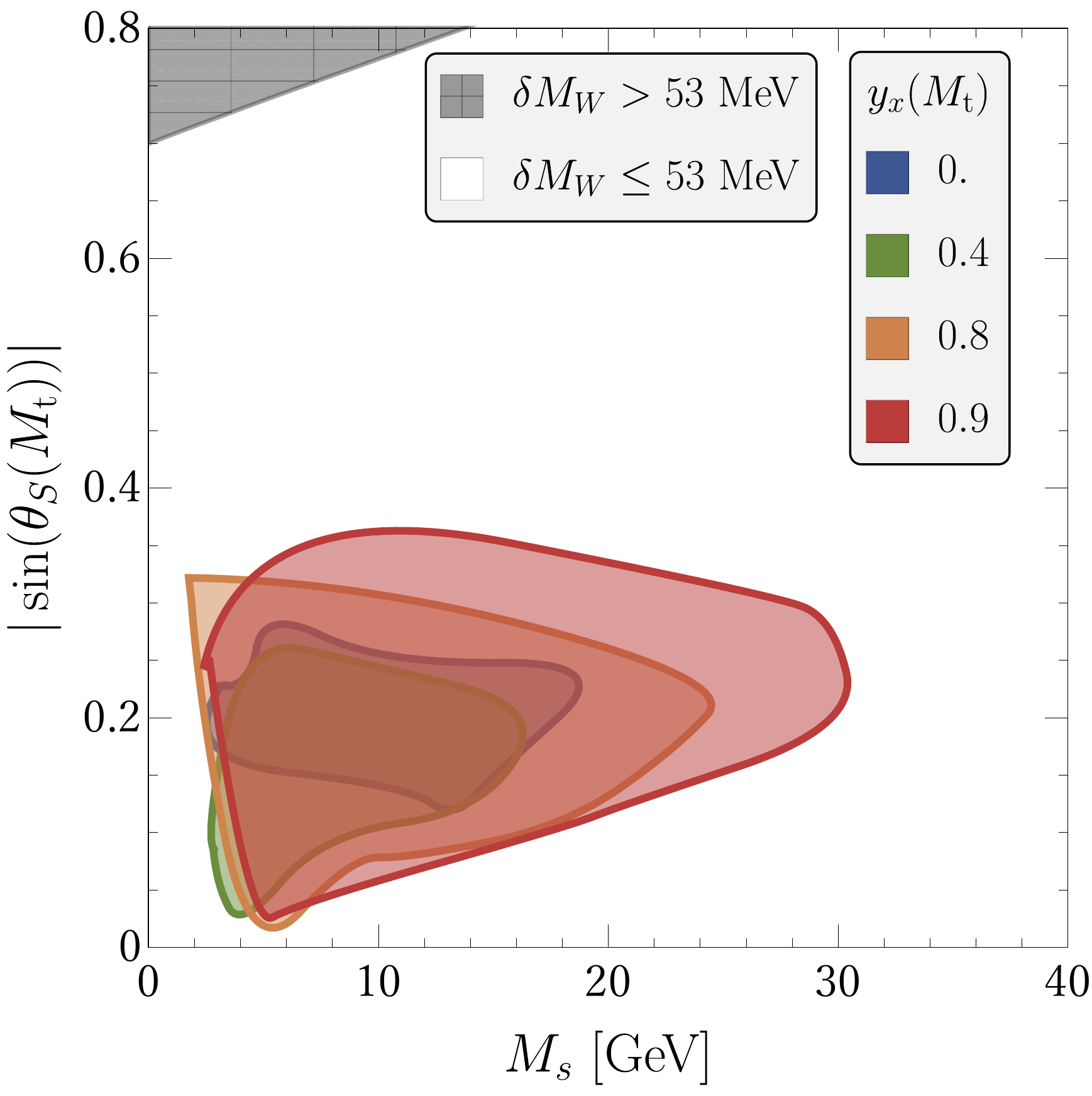}
\includegraphics[width=0.48\linewidth]{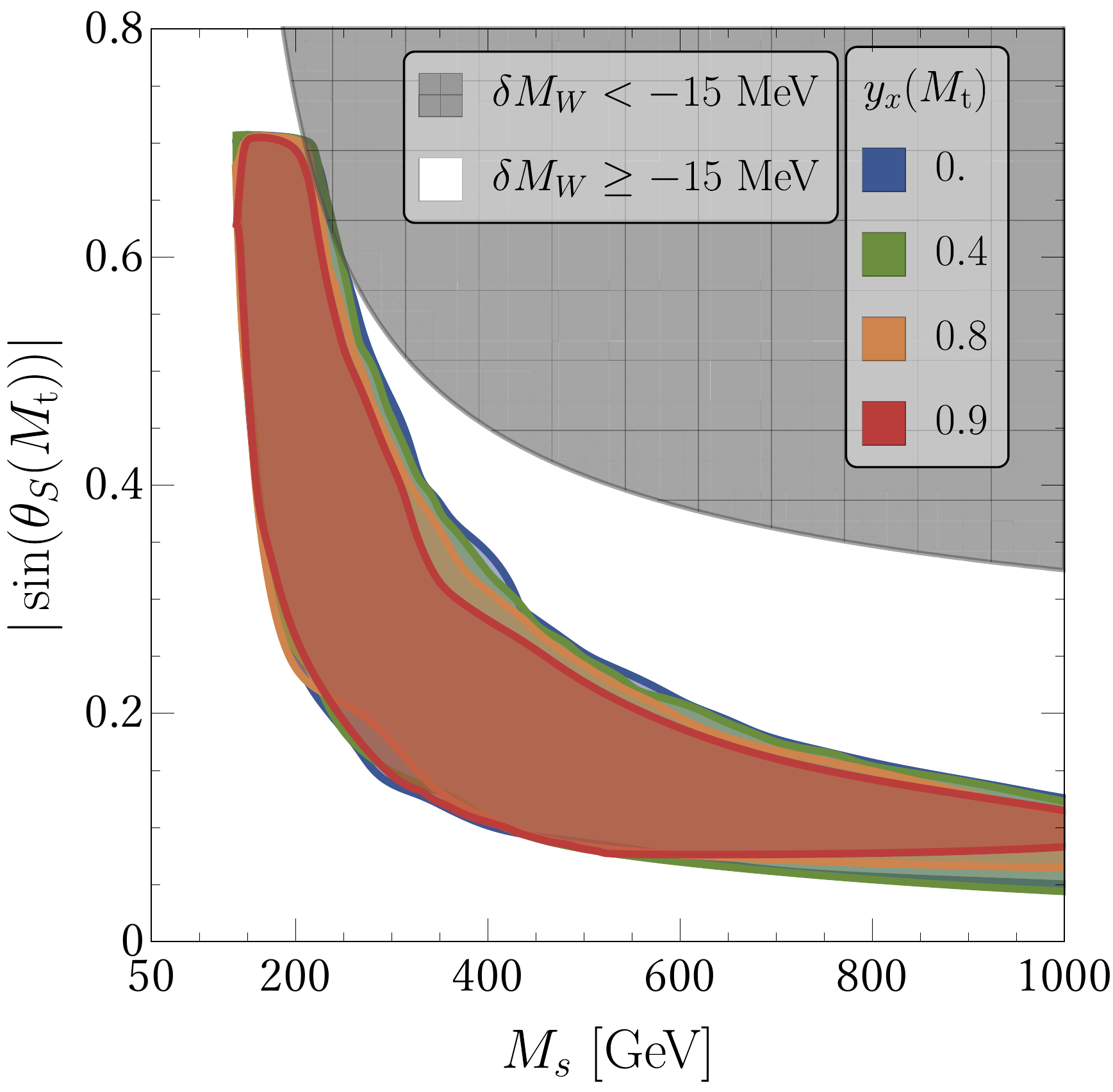}
\caption{\label{fig:final_result} 
Allowed parameter space in the $\msp-|\sin(\Theta_S)|$ 
plane at representative values of $y_x$ at two-loop accuracy. The region 
with a gray grid represents the portion of the parameter space where
$\delta\mwp$, i.e. the radiative corrections in the SWSM to the 
$W$-boson mass exceed the value given in the plot legend. 
}
\end{figure}
\begin{figure}[t]
\includegraphics[width=0.48\linewidth]{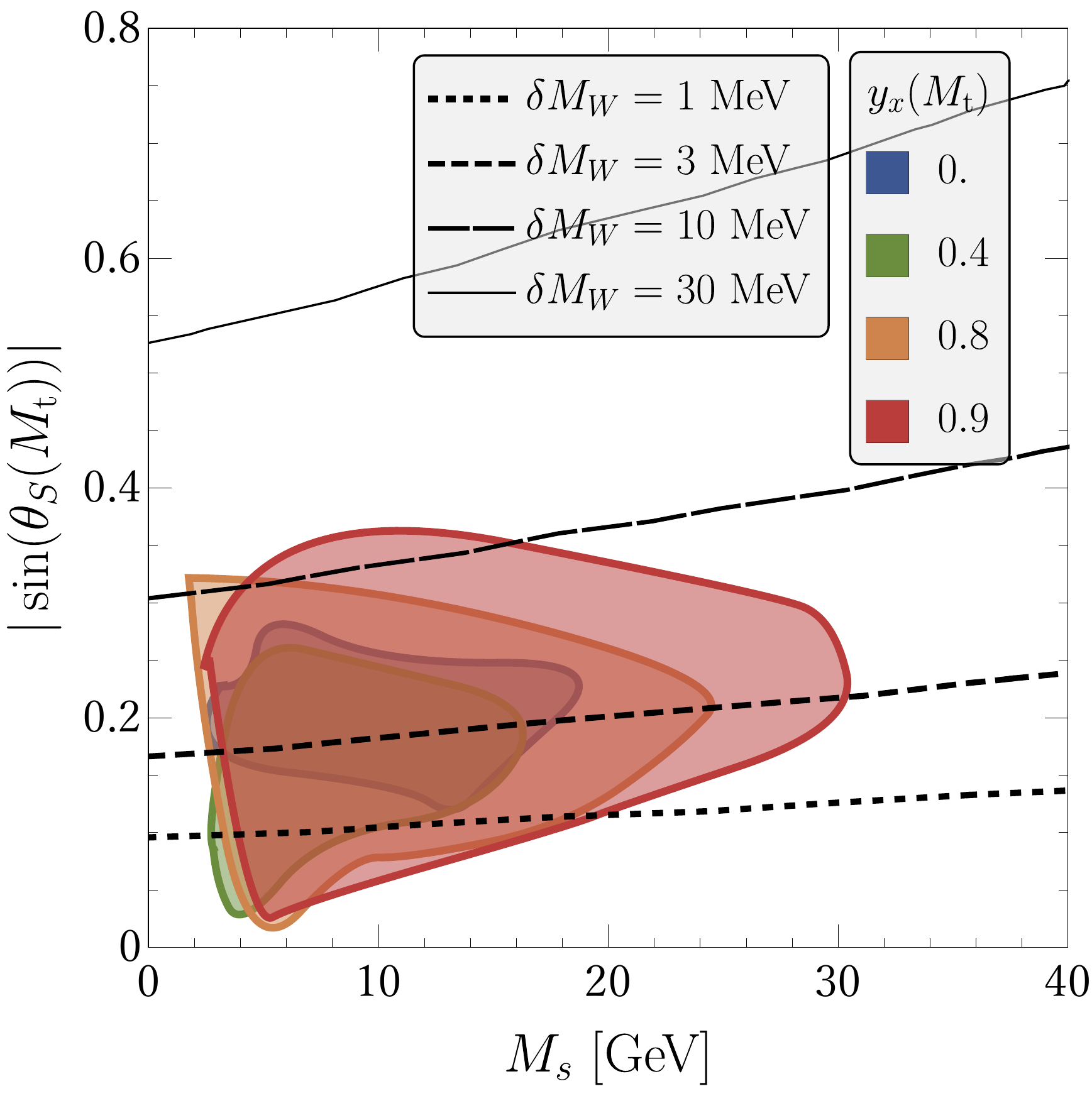}
\caption{\label{fig:final_result-CDF} 
Allowed parameter space in the $\msp-|\sin(\Theta_S)|$ 
plane at representative values of $y_x$ at two-loop accuracy for $\msp > \mhp$. 
The contours at selective values of $\delta\mwp$ represent the 
borderline in the parameter space below which the new scalar cannot 
be solely responsible to the increase of the $W$ boson mass by $\delta\mwp$
with respect to the SM value.}
\end{figure}

We computed the SW contributions $\delta\mwp^{\rm SW}$ to $M_W^{\text{theo.}}$ 
at one loop accuracy. We found, that the contribution of the new gauge sector 
is heavily suppressed due to the required smallness of the new gauge coupling 
$g_z \lesssim 10^{-4}$. The sterile neutrinos may change 
the measured value of the Fermi coupling $G_\rF$ affecting the mass of the $W$ 
boson already at tree level \cite{Fernandez-Martinez:2016lgt}. 
As a matter of fact right-handed neutrinos can provide significant contribution
to the $W$ boson mass \cite{Blennow:2022yfm}, although at the price of introducing
some tension with universality bounds. Hence, a proper account of the 
effect of sterile neutrinos is certainly warranted, but it is beyond the 
scope of the present paper and we leave it for a planned global scan of 
the parameter space. The contribution of the new scalar sector to $\mwp$ however, 
can be comparable to $\Delta\mwp$ \cite{Lopez-Val:2014jva,Robens:2015gla}. 
We present the SW correction $\delta\mwp$ in \eqref{app:1loopwmass} of
App.~\ref{app:electroweak-correction}, expressed with two free parameters, 
$\msp$ and $\sin^2\tS$. For the case of light new scalar, $\msp<\mhp$, the SW
correction is positve, while for the heavy one it is negative.   
The excluded regions obtained by (a) $\delta\mwp > 53$\,MeV and 
(b) $\delta\mwp < -15$\,MeV are presented in Fig.~\ref{fig:final_result} 
overlayed the region where the new scalar particle is allowed by vacuum 
stability, perturbativity and precision measurement of the Higgs boson mass. 
We see that the $W$-mass measurement at present uncertainties does not provide 
significant reduction of the parameter space. However, if the improved 
measurement published recently by the CDF collaboration \cite{CDF:2022hxs} 
will be confirmed, the stability of the vacuum in the high-mass region becomes
incompatible with the CDF-II $W$ mass as in that case the SW correction to the 
SM value is negative. The low-mass 
region also becomes significantly constrained.
In Fig.~\ref{fig:final_result-CDF} we present the allowed parameter space together
with contour lines representing the border of the excluded parameter space (below
the line) assuming a $\delta\mwp$ increase of the $W$ mass by selected 
benchmark values due to the new scalar in the self energy loop. Clearly, 
the large positive shift required to explain the CDF-II result is not 
compatible with the conditions of stability and perturbativity of the scalar 
sector of the model.

\section{Conclusions and outlook}

In this paper we have scanned the parameter space of the superweak extension of the
standard model in order to find the allowed parameter space of the scalar sector where 
the following assumptions are fulfilled: (i) the vacuum be stable and 
(ii) the model parameters remain perturbative up to the Planck scale, 
(iii) the pole mass of the Higgs boson must fall into its experimentally measured 
range. The first two of these constraints were taken into account in our preliminary 
work \cite{Peli:2019vtp}. In this paper we superseed that former study by taking 
into account the two-loop corrections both in the renormalization group equations 
of the running couplings as well as the measured value of the mass of the Higgs 
boson. We have taken into account the largest neutrino Yukawa coupling
$y_x$. In the limit of vanishing $y_x$ and neglecting the superweak gauge 
coupling, the model essentially reduces to the case of singlet scalar extension of 
the standard model.

In the two-loop analysis we found a non-empty region in the $\msp-\sin\tS$ parameter 
space for $\msp<\mhp$, increasing with $y_x$ up to $y_x(\mtp) \simeq 1$ where 
the condition of stability is not fulfilled any longer. Such a region have been
missed in the case of $y_x=0$ in earlier analyises of the singlet scalar extension
performed only at one-loop accuracy (see e.g.~\cite{Falkowski:2015iwa}).

Of course, there is a lot of experimental results that also constrain the parameter
space. The new physics contributions to electroweak precision observables as well 
as direct searches for the decay of a scalar particle into standard model ones 
provide strong constraints. Of those, we have studied only the effect of the 
experimental result on the mass of the $W$ boson in this paper. We saw that 
while $\mwp$ can indeed limit the parameter space, the current world average 
without the new CDF-II result cannot provide further constraint on the parameter 
space. If we also include the CDF II result in the average -- in spite of being 
incompatible with the previous average --, then the parameter space allowed 
by our assumptions become incompatible with the $W$-mass constraint. 
Clearly, it is of utmost importance to take into account all the available 
experimental constraints, not only from collider experiments, also from 
neutrino experiments. Such a complete study of the parameter space is beyond 
the scope of the present paper and we leave it to an upcoming study where we
plan to use the analytic expressions of the present work.

\section*{Acknowledgments}
We are grateful to members of the ELTE PPPhenogroup (pppheno.elte.hu/people),
especially to Josu Hernandez-Garcia for useful discussions.
This work was supported by grant K 125105 of the National Research,
Development and Innovation Fund in Hungary.  

\appendix
\section{Loop corrections to the scalar masses in the SWSM}
\label{app:oneloop}

On one hand, the quantum effects correct the tadpole equations \eqref{eq:tadpole}, such that
\begin{equation}\label{eq:tadpole-1loop}
\bsp
   0 &=
  v\biggl( -\mu_\phi^2 + \frac{1}{2}\lambda w^2 + \lambda_\phi v^2\biggr) + T_H,
  \\
   0 &=
  w\biggl( -\mu_\chi^2 + \frac{1}{2}\lambda v^2 + \lambda_\chi w^2\biggr) + T_S
  \,,
 \esp
\end{equation}
where $T_{\varphi}$ is the sum of all 1PI one-point functions with external leg 
$\varphi = H$ or $S$. On the other hand, the scalar self-energies 
$\Pi_{\varphi_I \varphi_J}$ -- i.e. the sum of all 1PI Feynmann graphs with 
external legs $\varphi_I$ and $\varphi_J$ -- correct directly the propagator matrix 
of the scalar fields $H$ and $S$. The inverse-propagator matrix after applying 
the tadpole equations \eqref{eq:tadpole-1loop} to eliminate the mass parameters 
$\mu_\phi$ and $\mu_\chi$ is given as
\begin{equation}\label{eq:propmatrix}
\begin{pmatrix}
    p^2  & 0\\
    0    & p^2
    \end{pmatrix}
-
\begin{pmatrix}
    2\lambda_\phi v^2 +  \tilde{\Pi}_{HH}(p^2) & \lambda v w + \Pi_{SH}(p^2)\\
    \lambda v w +  \Pi_{HS}(p^2)   &  2\lambda_\chi w^2 +\tilde{\Pi}_{SS}(p^2)
    \end{pmatrix}
,
\end{equation}
where $\tilde{\Pi}_{\varphi\varphi}(p^2) = \Pi_{\varphi\varphi}(p^2)-T_\varphi/\langle\varphi\rangle$ 
with $\langle\varphi\rangle = v$, $w$. In order to obtain the mixing angle $\tS$
\eqref{eq:thetas_1loop} and the scalar pole masses \eqref{eq:mh_1loop} and 
\eqref{eq:ms_1loop}, one has to diagonalize the real part of \eqref{eq:propmatrix},
leading to Eqs.~\eqref{eq:thetas_1loop}, \eqref{eq:mh_1loop} and 
\eqref{eq:ms_1loop}.

We have implemented a \texttt{SARAH} model file for the SWSM \cite{Iwamoto:2021wko},
and used it to compute the one-loop scalar self-energies and tadpoles in Feynman 
gauge ($\xi=1$). In the following, we list explicitly the one-loop corrections
to the scalar inverse-propagator matrix \eqref{eq:propmatrix}, after some 
simplifications of the \texttt{SARAH} output.

The Higgs self energy is
\begin{equation}
\bsp
\label{eq:pihh}
\tilde{\Pi}_{HH}(p^2) = &\kappa\biggl\{
  3 y_t^2\bigl(4 m_t^2 - p^2\bigr)B_0\bigl(p^2,m_t^2,m_t^2\bigr)
  -\frac{1}{2}\lambda^2 v^2 B_0\bigl(p^2,0,0\bigr) 
\\ &
  + \frac{\lambda w \sin(2\tS)}{2v}\biggl(A_0(m_s^2)-A_0(m_h^2)\biggr)
\\ &
  -\frac{1}{2}\bigl(6 \lambda_\phi v \cos^2\tS-  \lambda w \sin(2\tS) + \lambda v \sin^2 \tS\bigr)^2 B_0\bigl(p^2,m_h^2,m_h^2\bigr)
\\ &
  -\frac{1}{2}\bigl(\lambda v \cos^2 \tS+  \lambda w \sin(2\tS) + 6 \lambda_\phi v \sin^2\tS\bigr)^2 B_0\bigl(p^2,m_s^2,m_s^2\bigr)
\\ &
  -\frac{1}{2}\bigl(\lambda w \cos(2\tS) + (3 \lambda_\phi -\lambda/2) v \sin(2\tS)\bigr)^2 B_0\bigl(p^2,m_s^2,m_h^2\bigr)
\\ &
   +\frac{1}{v^2}\biggl(
   \bigl(4 p^2 m_W^2 - 12 m_W^4- 4 \lambda_\phi^2 v^4\bigr)B_0\bigl(p^2,m_W^2,m_W^2\bigr) +8 m_W^2
   \biggr)
\\ &
   +\frac{1}{2 v^2}\biggl(
   \bigl(4 p^2 m_Z^2 - 12 m_Z^4- 4 \lambda_\phi^2 v^4\bigr)B_0\bigl(p^2,m_Z^2,m_Z^2\bigr) +8 m_Z^2
   \biggr)
  \biggr\}
\esp
\end{equation}
where $\kappa=(4\pi)^{-2}$ and $m_p(\mu)$ is the running mass of particle $p$ as computed 
using tree level relations. If the elements of the Dirac neutrino mass matrix 
$\mathbf{M}_D$ are much smaller than those of the Majorana neutrino mass matrix
$\mathbf{M}_N$, i.e.~$\bigl(\mathbf{M}_D\bigr)_{ij} \ll \bigl(\mathbf{M}_N\bigr)_{kk}$ 
for any $i,j,k \in (1,2,3)$, then the loop correction to the scalar mass from the 
active neutrinos is negligible and the sterile neutrinos contribute to 
$\tilde{\Pi}_{SS}$ only, namely
\begin{equation}
\bsp
\label{eq:piss}
\tilde{\Pi}_{SS}(p^2)= 
\kappa &\biggl\{
\frac{1}{2}\sum_{i=1}^3y_{x,i}^2\bigl(4 m_{N,i}^2 -p^2\bigr)B_0\bigl(p^2,m_{N,i}^2,m_{N,i}^2 \bigr)
\\ &
+ \frac{\lambda v \sin(2\tS)}{2w}\biggl(A_0(m_s^2)-A_0(m_h^2)\biggr)
\\  &
-\frac{1}{2}\bigl(\lambda w \cos^2\tS-  \lambda v \sin(2\tS) + 6 \lambda_\chi w \sin^2 \tS\bigr)^2 B_0\bigl(p^2,m_h^2,m_h^2\bigr)
\\  &
-\frac{1}{2}\bigl(6 \lambda_\chi w \cos^2 \tS+  \lambda v \sin(2\tS) +  \lambda w \sin^2\tS\bigr)^2 B_0\bigl(p^2,m_s^2,m_s^2\bigr)
\\  &\quad
-\frac{1}{2}\bigl(\lambda v \cos(2\tS)  - (3 \lambda_\chi -\lambda/2) w \sin(2\tS)\bigr)^2 B_0\bigl(p^2,m_s^2,m_h^2\bigr)
\\  &  
-w^2 \biggl(\lambda^2 B_0\bigl(p^2,m_{W}^2,m_{W}^2\bigr)
+ \frac{1}{2}\lambda^2 B_0\bigl(p^2,m_{Z}^2,m_{Z}^2\bigr) 
- 2 \lambda_\chi^2   B_0\bigl(p^2,0,0\bigr) \biggr)
\biggr\}
\esp
\end{equation}
and
\begin{equation}
\bsp
\label{eq:pihs}
\Pi_{HS}&(p^2) =  \Pi_{SH}(p^2) =
\\ &
= \kappa  \biggl\{
  \frac{1}{2}\lambda  \sin(2\tS)\biggl(A_0(m_h^2)-A_0(m_s^2)\biggr)
\\ &
  +\frac{1}{2}\bigl(6\lambda_\phi v \cos^2\tS-  \lambda w \sin(2\tS) +\lambda v \sin^2 \tS\bigr)
\\ &\qquad
  \times\bigl(\lambda w \cos^2\tS-  \lambda v \sin(2\tS) + 6 \lambda_\chi w \sin^2 \tS\bigr)
  B_0\bigl(p^2,m_h^2,m_h^2\bigr)
\\ &
  -\frac{1}{2}\bigl(6 \lambda_\chi w \cos^2 \tS+  \lambda v \sin(2\tS) +  \lambda w \sin^2\tS\bigr)
\\ &\qquad
  \times  
  \bigl(\lambda v \cos^2 \tS+  \lambda w \sin(2\tS) + 6 \lambda_\phi w \sin^2\tS\bigr)
  B_0\bigl(p^2,m_s^2,m_s^2\bigr)
  \biggr]
\\ &
  -\frac{1}{4}\bigl(2\lambda v \cos(2\tS)  - (6 \lambda_\chi -\lambda) w \sin(2\tS)\bigr)
\\ &\qquad
  \times  
   \bigl(2\lambda w \cos(2\tS) + (6 \lambda_\phi -\lambda) v \sin(2\tS)\bigr)  
  B_0\bigl(p^2,m_s^2,m_h^2\bigr)
\\ &  
  -\lambda v w \biggl( 
  2\lambda_\phi B_0\bigl(p^2,M_{W}^2,M_{W}^2\bigr)
  + \lambda_\phi B_0\bigl(p^2,M_{Z}^2,M_{Z}^2\bigr)
  + \lambda_\chi B_0\bigl(p^2,M_{Z'}^2,M_{Z'}^2\bigr) \biggr)
  \biggr\}
  \,.
\esp
\end{equation}
We have neglected terms proportional to $g_z$ and the mass $M_Z'$ of the $Z'$ boson
as $M_{Z'} \ll v,\, w$. 
Each coupling and masses in the one-loop contributions (\ref{eq:pihh}), (\ref{eq:piss}) and
(\ref{eq:pihs}) is the running parameter depending on the renormalization scale $\mu$, 
and the vacuum expectation values $v$ and $w$ are the gauge and scale dependent 
running VEVs.  We suppressed the scale dependence for easier reading. 
The masses $m_x$ correspond to the tree-level formulae, but with 
running couplings
\begin{equation}
    m_t = \frac{1}{\sqrt{2}}y_t v\,, \quad
    m_W = \frac{1}{2}g_\rL v\,, \quad
    m_Z = \frac{1}{2}\sqrt{g_Y^2+g_\rL^2}~v\,, \quad
    \text{and}\quad
     m_{N,i} = \frac{y_{x,i}}{\sqrt{2}} w\,.  
\end{equation}
and $\tS$, $m_h$, $m_s$ and  correspond to Eqs.~(\ref{eq:thetas_tree}), (\ref{eq:mh_tree})
and  (\ref{eq:ms_tree}). The loop functions $A_0$ and $B_0$ are given as 
\begin{eqnarray}
A_0(m^2) &=& m^2 \biggl( 1 - \ln\biggl(\frac{m^2}{\mu^2}\biggr)\biggr),
\\
B_0(s,m_1^2,m_2^2) &=& 
-\int_0^1\!\rd u\ln\biggl(\frac{u\,m_1^2 + (1-u) m_2^2-u(1-u)s}{\mu^2}\biggr) .
\end{eqnarray}
In the special case of vanishing masses, the latter reduces to
\begin{equation}
B_0(s,0,0) = 2 - \ln\biggl(-\frac{s}{\mu^2}\biggr) = 2- \ln\biggl|\frac{s}{\mu^2}\biggr| +\ri\pi \Theta(s)
\end{equation}
where $\Theta$ is the Heaviside step function. 
In this work, we are not concerned with the decay width $\Gamma_{h/s}$ of 
the scalar bosons. The imaginary parts of the one-loop self energies 
and tadpoles contribute to the decay widths, and we neglect those completely.

\section{One-loop corrections to the gauge bosons and electroweak input parameters in the SWSM}\label{app:electroweak-correction}

The SWSM introduces new corrections to the $W$ and $Z$ gauge boson 
self-energies.
The radiative corrections from the new gauge sector are neglected due to coupling suppression $g_z \lesssim 10^{-4}$, whereas the sterile neutrinos may not only contribute radiatively through the PMNS matrix but also contribute at tree level by affecting the Fermi coupling $G_\text{F}$ through the low energy muon decay. We neglect the neutrino contributions (to be investigated in an upcoming paper) and focus here on the pure scalar radiative corrections.
In the $\overline{\text{MS}}$ 
scheme 
the scalar SW contribution to the gauge boson self-energies is
\begin{equation}\label{eq:wz_selfenergy}
\Pi^{\text{SW}}_{VV}(p^2) = \frac{\sin^2\tS}{16\pi^2}\frac{m_V^2}{v^2}
\biggl( F(p^2,m_V^2,m_s^2)-F(p^2,m_V^2,m_h^2) \biggr),
\quad V=W,\:Z
\end{equation}
where the loop function $F$ is defined as
\begin{equation}
\bsp
F(s,m_1^2,m_2^2) &= \frac{2}{3}\biggl(m_1^2 + m_2^2-\frac{s}{3}\biggr)
\\&
+\biggl(\frac{s + m_1^2 - m_2^2}{3s}\biggr)
\Big(A_0(m_1^2) - A_0(m_2^2)\Big) - \frac{1}{3} A_0(m_2^2)
\\&
-\frac{1}{3}\biggl(\frac{(m_1^2 - m_2^2)^2}{s} - 2(m_2^2-5 m_1^2) + s\biggr) B_0(s,m_1^2,m_2^2).
\esp
\end{equation}
We have checked in the $R_\xi$ gauge, that the scalar contribution
$\Pi^{\text{SW}}_{VV}(p^2)$ is explicitly independent of the gauge
parameter $\xi$, hence gauge invariant. Furthermore, 
$\Pi^{\text{SW}}_{VV}(M_V^2)$ is independent of the the 
renormalization scale at one loop accuracy.

The shift in the electroweak input parameters due to the SW 
corrections is then 
\begin{eqnarray}
    \frac{\delta g_\rL}{g_\rL} &=& \frac{1}{4 m_W^2 - 2 m_Z^2}\biggl(\Pi^{\text{SW}}_{WW}(0)- \frac{m_W^2}{m_Z^2}\Pi^{\text{SW}}_{ZZ}(\mzp^2)\biggr),
    \\
    \frac{\delta g_Y}{g_Y} &=& \frac{m_Z^2-m_W^2}{4 m_W^2 - 2 m_Z^2}\biggl(-\frac{\Pi^{\text{SW}}_{WW}(0)}{m_W^2}+ \frac{\Pi^{\text{SW}}_{ZZ}(\mzp^2)}{m_Z^2}\biggr),  
    \\
    \frac{\delta v}{v} &=& -\frac{\Pi^{\text{SW}}_{WW}(0)}{2 m_W^2},
\end{eqnarray}
which agrees with Eq.~(22) of \cite{Falkowski:2015iwa}.
The $W$ boson pole mass is then given as 
\begin{equation}
    \mwp^2 = \frac{1}{4}(g_\rL+\delta g_\rL)^2(v+\delta v)^2 + \Pi^{\text{SM}}_{WW}(\mwp^2) + \Pi^{\text{SW}}_{WW}(\mwp^2).
\end{equation}
It is convenient to express $\mwp$ as the sum of the SM $\mwp^{\text{theo.}}$ and the new physics contribution $\delta M_W$ as
\begin{equation}\label{app:1loopwmass}
    \mwp = \mwp^{\text{theo.}} + \delta M_W,
    \quad\text{with}\quad
      \delta M_W = \mwp \frac{\delta v}{v} + \frac{1}{2}\delta g_\rL v + \frac{\Pi^{\text{SW}}_{WW}(\mwp)}{2\mwp}.
\end{equation}


\begin{thebibliography}{42}%
\makeatletter
\providecommand \@ifxundefined [1]{%
 \@ifx{#1\undefined}
}%
\providecommand \@ifnum [1]{%
 \ifnum #1\expandafter \@firstoftwo
 \else \expandafter \@secondoftwo
 \fi
}%
\providecommand \@ifx [1]{%
 \ifx #1\expandafter \@firstoftwo
 \else \expandafter \@secondoftwo
 \fi
}%
\providecommand \natexlab [1]{#1}%
\providecommand \enquote  [1]{``#1''}%
\providecommand \bibnamefont  [1]{#1}%
\providecommand \bibfnamefont [1]{#1}%
\providecommand \citenamefont [1]{#1}%
\providecommand \href@noop [0]{\@secondoftwo}%
\providecommand \href [0]{\begingroup \@sanitize@url \@href}%
\providecommand \@href[1]{\@@startlink{#1}\@@href}%
\providecommand \@@href[1]{\endgroup#1\@@endlink}%
\providecommand \@sanitize@url [0]{\catcode `\\12\catcode `\$12\catcode
  `\&12\catcode `\#12\catcode `\^12\catcode `\_12\catcode `\%12\relax}%
\providecommand \@@startlink[1]{}%
\providecommand \@@endlink[0]{}%
\providecommand \url  [0]{\begingroup\@sanitize@url \@url }%
\providecommand \@url [1]{\endgroup\@href {#1}{\urlprefix }}%
\providecommand \urlprefix  [0]{URL }%
\providecommand \Eprint [0]{\href }%
\providecommand \doibase [0]{https://doi.org/}%
\providecommand \selectlanguage [0]{\@gobble}%
\providecommand \bibinfo  [0]{\@secondoftwo}%
\providecommand \bibfield  [0]{\@secondoftwo}%
\providecommand \translation [1]{[#1]}%
\providecommand \BibitemOpen [0]{}%
\providecommand \bibitemStop [0]{}%
\providecommand \bibitemNoStop [0]{.\EOS\space}%
\providecommand \EOS [0]{\spacefactor3000\relax}%
\providecommand \BibitemShut  [1]{\csname bibitem#1\endcsname}%
\let\auto@bib@innerbib\@empty
\bibitem [{\citenamefont {Fukuda}\ \emph {et~al.}(1998)\citenamefont {Fukuda}
  \emph {et~al.}}]{Fukuda:1998mi}%
  \BibitemOpen
  \bibfield  {author} {\bibinfo {author} {\bibfnamefont {Y.}~\bibnamefont
  {Fukuda}} \emph {et~al.} (\bibinfo {collaboration} {Super-Kamiokande}),\
  }\bibfield  {title} {\bibinfo {title} {{Evidence for oscillation of
  atmospheric neutrinos}},\ }\href
  {https://doi.org/10.1103/PhysRevLett.81.1562} {\bibfield  {journal} {\bibinfo
   {journal} {Phys. Rev. Lett.}\ }\textbf {\bibinfo {volume} {81}},\ \bibinfo
  {pages} {1562} (\bibinfo {year} {1998})},\ \Eprint
  {https://arxiv.org/abs/hep-ex/9807003} {arXiv:hep-ex/9807003 [hep-ex]}
  \BibitemShut {NoStop}%
\bibitem [{\citenamefont {Ahmad}\ \emph {et~al.}(2001)\citenamefont {Ahmad}
  \emph {et~al.}}]{Ahmad:2001an}%
  \BibitemOpen
  \bibfield  {author} {\bibinfo {author} {\bibfnamefont {Q.~R.}\ \bibnamefont
  {Ahmad}} \emph {et~al.} (\bibinfo {collaboration} {SNO Collaboration}),\
  }\bibfield  {title} {\bibinfo {title} {Measurement of the rate of
  ${\nu}_{e}+d \rightarrow p+p+e^-$ interactions produced by $^{8}B$ solar
  neutrinos at the sudbury neutrino observatory},\ }\href
  {https://doi.org/10.1103/PhysRevLett.87.071301} {\bibfield  {journal}
  {\bibinfo  {journal} {Phys. Rev. Lett.}\ }\textbf {\bibinfo {volume} {87}},\
  \bibinfo {pages} {071301} (\bibinfo {year} {2001})}\BibitemShut {NoStop}%
\bibitem [{\citenamefont {Bezrukov}\ \emph {et~al.}(2012)\citenamefont
  {Bezrukov}, \citenamefont {Kalmykov}, \citenamefont {Kniehl},\ and\
  \citenamefont {Shaposhnikov}}]{Bezrukov:2012sa}%
  \BibitemOpen
  \bibfield  {author} {\bibinfo {author} {\bibfnamefont {F.}~\bibnamefont
  {Bezrukov}}, \bibinfo {author} {\bibfnamefont {M.~Y.}\ \bibnamefont
  {Kalmykov}}, \bibinfo {author} {\bibfnamefont {B.~A.}\ \bibnamefont
  {Kniehl}},\ and\ \bibinfo {author} {\bibfnamefont {M.}~\bibnamefont
  {Shaposhnikov}},\ }\bibfield  {title} {\bibinfo {title} {{Higgs Boson Mass
  and New Physics}},\ }\href {https://doi.org/10.1007/JHEP10(2012)140}
  {\bibfield  {journal} {\bibinfo  {journal} {JHEP}\ }\textbf {\bibinfo
  {volume} {10}},\ \bibinfo {pages} {140}},\ \Eprint
  {https://arxiv.org/abs/1205.2893} {arXiv:1205.2893 [hep-ph]} \BibitemShut
  {NoStop}%
\bibitem [{\citenamefont {Degrassi}\ \emph {et~al.}(2012)\citenamefont
  {Degrassi}, \citenamefont {Di~Vita}, \citenamefont {Elias-Miro},
  \citenamefont {Espinosa}, \citenamefont {Giudice}, \citenamefont {Isidori},\
  and\ \citenamefont {Strumia}}]{Degrassi:2012ry}%
  \BibitemOpen
  \bibfield  {author} {\bibinfo {author} {\bibfnamefont {G.}~\bibnamefont
  {Degrassi}}, \bibinfo {author} {\bibfnamefont {S.}~\bibnamefont {Di~Vita}},
  \bibinfo {author} {\bibfnamefont {J.}~\bibnamefont {Elias-Miro}}, \bibinfo
  {author} {\bibfnamefont {J.~R.}\ \bibnamefont {Espinosa}}, \bibinfo {author}
  {\bibfnamefont {G.~F.}\ \bibnamefont {Giudice}}, \bibinfo {author}
  {\bibfnamefont {G.}~\bibnamefont {Isidori}},\ and\ \bibinfo {author}
  {\bibfnamefont {A.}~\bibnamefont {Strumia}},\ }\bibfield  {title} {\bibinfo
  {title} {{Higgs mass and vacuum stability in the Standard Model at NNLO}},\
  }\href {https://doi.org/10.1007/JHEP08(2012)098} {\bibfield  {journal}
  {\bibinfo  {journal} {JHEP}\ }\textbf {\bibinfo {volume} {08}},\ \bibinfo
  {pages} {098}},\ \Eprint {https://arxiv.org/abs/1205.6497} {arXiv:1205.6497
  [hep-ph]} \BibitemShut {NoStop}%
\bibitem [{\citenamefont {Hinshaw}\ \emph {et~al.}(2013)\citenamefont {Hinshaw}
  \emph {et~al.}}]{Hinshaw:2012aka}%
  \BibitemOpen
  \bibfield  {author} {\bibinfo {author} {\bibfnamefont {G.}~\bibnamefont
  {Hinshaw}} \emph {et~al.} (\bibinfo {collaboration} {WMAP}),\ }\bibfield
  {title} {\bibinfo {title} {{Nine-Year Wilkinson Microwave Anisotropy Probe
  (WMAP) Observations: Cosmological Parameter Results}},\ }\href
  {https://doi.org/10.1088/0067-0049/208/2/19} {\bibfield  {journal} {\bibinfo
  {journal} {Astrophys. J. Suppl.}\ }\textbf {\bibinfo {volume} {208}},\
  \bibinfo {pages} {19} (\bibinfo {year} {2013})},\ \Eprint
  {https://arxiv.org/abs/1212.5226} {arXiv:1212.5226 [astro-ph.CO]}
  \BibitemShut {NoStop}%
\bibitem [{\citenamefont {Aghanim}\ \emph {et~al.}(2020)\citenamefont {Aghanim}
  \emph {et~al.}}]{Aghanim:2018eyx}%
  \BibitemOpen
  \bibfield  {author} {\bibinfo {author} {\bibfnamefont {N.}~\bibnamefont
  {Aghanim}} \emph {et~al.} (\bibinfo {collaboration} {Planck}),\ }\bibfield
  {title} {\bibinfo {title} {{Planck 2018 results. VI. Cosmological
  parameters}},\ }\href {https://doi.org/10.1051/0004-6361/201833910}
  {\bibfield  {journal} {\bibinfo  {journal} {Astron. Astrophys.}\ }\textbf
  {\bibinfo {volume} {641}},\ \bibinfo {pages} {A6} (\bibinfo {year} {2020})},\
  \Eprint {https://arxiv.org/abs/1807.06209} {arXiv:1807.06209 [astro-ph.CO]}
  \BibitemShut {NoStop}%
\bibitem [{\citenamefont {Eisenstein}\ \emph {et~al.}(2005)\citenamefont
  {Eisenstein} \emph {et~al.}}]{Eisenstein:2005su}%
  \BibitemOpen
  \bibfield  {author} {\bibinfo {author} {\bibfnamefont {D.~J.}\ \bibnamefont
  {Eisenstein}} \emph {et~al.} (\bibinfo {collaboration} {SDSS}),\ }\bibfield
  {title} {\bibinfo {title} {{Detection of the Baryon Acoustic Peak in the
  Large-Scale Correlation Function of SDSS Luminous Red Galaxies}},\ }\href
  {https://doi.org/10.1086/466512} {\bibfield  {journal} {\bibinfo  {journal}
  {Astrophys. J.}\ }\textbf {\bibinfo {volume} {633}},\ \bibinfo {pages} {560}
  (\bibinfo {year} {2005})},\ \Eprint {https://arxiv.org/abs/astro-ph/0501171}
  {arXiv:astro-ph/0501171} \BibitemShut {NoStop}%
\bibitem [{\citenamefont {Sofue}\ and\ \citenamefont
  {Rubin}(2001)}]{Sofue:2000jx}%
  \BibitemOpen
  \bibfield  {author} {\bibinfo {author} {\bibfnamefont {Y.}~\bibnamefont
  {Sofue}}\ and\ \bibinfo {author} {\bibfnamefont {V.}~\bibnamefont {Rubin}},\
  }\bibfield  {title} {\bibinfo {title} {{Rotation curves of spiral
  galaxies}},\ }\href {https://doi.org/10.1146/annurev.astro.39.1.137}
  {\bibfield  {journal} {\bibinfo  {journal} {Ann. Rev. Astron. Astrophys.}\
  }\textbf {\bibinfo {volume} {39}},\ \bibinfo {pages} {137} (\bibinfo {year}
  {2001})},\ \Eprint {https://arxiv.org/abs/astro-ph/0010594}
  {arXiv:astro-ph/0010594} \BibitemShut {NoStop}%
\bibitem [{\citenamefont {Bartelmann}\ and\ \citenamefont
  {Schneider}(2001)}]{Bartelmann:1999yn}%
  \BibitemOpen
  \bibfield  {author} {\bibinfo {author} {\bibfnamefont {M.}~\bibnamefont
  {Bartelmann}}\ and\ \bibinfo {author} {\bibfnamefont {P.}~\bibnamefont
  {Schneider}},\ }\bibfield  {title} {\bibinfo {title} {{Weak gravitational
  lensing}},\ }\href {https://doi.org/10.1016/S0370-1573(00)00082-X} {\bibfield
   {journal} {\bibinfo  {journal} {Phys. Rept.}\ }\textbf {\bibinfo {volume}
  {340}},\ \bibinfo {pages} {291} (\bibinfo {year} {2001})},\ \Eprint
  {https://arxiv.org/abs/astro-ph/9912508} {arXiv:astro-ph/9912508}
  \BibitemShut {NoStop}%
\bibitem [{\citenamefont {Aoyama}\ \emph {et~al.}(2020)\citenamefont {Aoyama}
  \emph {et~al.}}]{Aoyama:2020ynm}%
  \BibitemOpen
  \bibfield  {author} {\bibinfo {author} {\bibfnamefont {T.}~\bibnamefont
  {Aoyama}} \emph {et~al.},\ }\bibfield  {title} {\bibinfo {title} {{The
  anomalous magnetic moment of the muon in the Standard Model}},\ }\href
  {https://doi.org/10.1016/j.physrep.2020.07.006} {\bibfield  {journal}
  {\bibinfo  {journal} {Phys. Rept.}\ }\textbf {\bibinfo {volume} {887}},\
  \bibinfo {pages} {1} (\bibinfo {year} {2020})},\ \Eprint
  {https://arxiv.org/abs/2006.04822} {arXiv:2006.04822 [hep-ph]} \BibitemShut
  {NoStop}%
\bibitem [{\citenamefont {Bennett}\ \emph {et~al.}(2006)\citenamefont {Bennett}
  \emph {et~al.}}]{Muong-2:2006rrc}%
  \BibitemOpen
  \bibfield  {author} {\bibinfo {author} {\bibfnamefont {G.~W.}\ \bibnamefont
  {Bennett}} \emph {et~al.} (\bibinfo {collaboration} {Muon g-2}),\ }\bibfield
  {title} {\bibinfo {title} {{Final Report of the Muon E821 Anomalous Magnetic
  Moment Measurement at BNL}},\ }\href
  {https://doi.org/10.1103/PhysRevD.73.072003} {\bibfield  {journal} {\bibinfo
  {journal} {Phys. Rev. D}\ }\textbf {\bibinfo {volume} {73}},\ \bibinfo
  {pages} {072003} (\bibinfo {year} {2006})},\ \Eprint
  {https://arxiv.org/abs/hep-ex/0602035} {arXiv:hep-ex/0602035} \BibitemShut
  {NoStop}%
\bibitem [{\citenamefont {Abi}\ \emph {et~al.}(2021)\citenamefont {Abi} \emph
  {et~al.}}]{Muong-2:2021ojo}%
  \BibitemOpen
  \bibfield  {author} {\bibinfo {author} {\bibfnamefont {B.}~\bibnamefont
  {Abi}} \emph {et~al.} (\bibinfo {collaboration} {Muon g-2}),\ }\bibfield
  {title} {\bibinfo {title} {{Measurement of the Positive Muon Anomalous
  Magnetic Moment to 0.46 ppm}},\ }\href
  {https://doi.org/10.1103/PhysRevLett.126.141801} {\bibfield  {journal}
  {\bibinfo  {journal} {Phys. Rev. Lett.}\ }\textbf {\bibinfo {volume} {126}},\
  \bibinfo {pages} {141801} (\bibinfo {year} {2021})},\ \Eprint
  {https://arxiv.org/abs/2104.03281} {arXiv:2104.03281 [hep-ex]} \BibitemShut
  {NoStop}%
\bibitem [{\citenamefont {Borsanyi}\ \emph {et~al.}(2021)\citenamefont
  {Borsanyi} \emph {et~al.}}]{Borsanyi:2020mff}%
  \BibitemOpen
  \bibfield  {author} {\bibinfo {author} {\bibfnamefont {S.}~\bibnamefont
  {Borsanyi}} \emph {et~al.},\ }\bibfield  {title} {\bibinfo {title} {{Leading
  hadronic contribution to the muon magnetic moment from lattice QCD}},\ }\href
  {https://doi.org/10.1038/s41586-021-03418-1} {\bibfield  {journal} {\bibinfo
  {journal} {Nature}\ }\textbf {\bibinfo {volume} {593}},\ \bibinfo {pages}
  {51} (\bibinfo {year} {2021})},\ \Eprint {https://arxiv.org/abs/2002.12347}
  {arXiv:2002.12347 [hep-lat]} \BibitemShut {NoStop}%
\bibitem [{\citenamefont {Holdom}(1986)}]{Holdom:1985ag}%
  \BibitemOpen
  \bibfield  {author} {\bibinfo {author} {\bibfnamefont {B.}~\bibnamefont
  {Holdom}},\ }\bibfield  {title} {\bibinfo {title} {{Two U(1)'s and $\epsilon$
  charge shifts}},\ }\href {https://doi.org/10.1016/0370-2693(86)91377-8}
  {\bibfield  {journal} {\bibinfo  {journal} {Phys. Lett. B}\ }\textbf
  {\bibinfo {volume} {166}},\ \bibinfo {pages} {196} (\bibinfo {year}
  {1986})}\BibitemShut {NoStop}%
\bibitem [{\citenamefont {Pospelov}\ \emph {et~al.}(2008)\citenamefont
  {Pospelov}, \citenamefont {Ritz},\ and\ \citenamefont
  {Voloshin}}]{Pospelov:2007mp}%
  \BibitemOpen
  \bibfield  {author} {\bibinfo {author} {\bibfnamefont {M.}~\bibnamefont
  {Pospelov}}, \bibinfo {author} {\bibfnamefont {A.}~\bibnamefont {Ritz}},\
  and\ \bibinfo {author} {\bibfnamefont {M.~B.}\ \bibnamefont {Voloshin}},\
  }\bibfield  {title} {\bibinfo {title} {{Secluded WIMP Dark Matter}},\ }\href
  {https://doi.org/10.1016/j.physletb.2008.02.052} {\bibfield  {journal}
  {\bibinfo  {journal} {Phys. Lett. B}\ }\textbf {\bibinfo {volume} {662}},\
  \bibinfo {pages} {53} (\bibinfo {year} {2008})},\ \Eprint
  {https://arxiv.org/abs/0711.4866} {arXiv:0711.4866 [hep-ph]} \BibitemShut
  {NoStop}%
\bibitem [{\citenamefont {Schabinger}\ and\ \citenamefont
  {Wells}(2005)}]{Schabinger:2005ei}%
  \BibitemOpen
  \bibfield  {author} {\bibinfo {author} {\bibfnamefont {R.~M.}\ \bibnamefont
  {Schabinger}}\ and\ \bibinfo {author} {\bibfnamefont {J.~D.}\ \bibnamefont
  {Wells}},\ }\bibfield  {title} {\bibinfo {title} {{A Minimal spontaneously
  broken hidden sector and its impact on Higgs boson physics at the large
  hadron collider}},\ }\href {https://doi.org/10.1103/PhysRevD.72.093007}
  {\bibfield  {journal} {\bibinfo  {journal} {Phys. Rev. D}\ }\textbf {\bibinfo
  {volume} {72}},\ \bibinfo {pages} {093007} (\bibinfo {year} {2005})},\
  \Eprint {https://arxiv.org/abs/0509209} {arXiv:0509209 [hep-ph]} \BibitemShut
  {NoStop}%
\bibitem [{\citenamefont {Patt}\ and\ \citenamefont
  {Wilczek}(2006)}]{Patt:2006fwx}%
  \BibitemOpen
  \bibfield  {author} {\bibinfo {author} {\bibfnamefont {B.}~\bibnamefont
  {Patt}}\ and\ \bibinfo {author} {\bibfnamefont {F.}~\bibnamefont {Wilczek}},\
  }\bibfield  {title} {\bibinfo {title} {{Higgs-field portal into hidden
  sectors}}\ }\href@noop {} {} (\bibinfo {year} {2006}),\ \Eprint
  {https://arxiv.org/abs/0605188} {arXiv:0605188 [hep-ph]} \BibitemShut
  {NoStop}%
\bibitem [{\citenamefont {Falkowski}\ \emph {et~al.}(2015)\citenamefont
  {Falkowski}, \citenamefont {Gross},\ and\ \citenamefont
  {Lebedev}}]{Falkowski:2015iwa}%
  \BibitemOpen
  \bibfield  {author} {\bibinfo {author} {\bibfnamefont {A.}~\bibnamefont
  {Falkowski}}, \bibinfo {author} {\bibfnamefont {C.}~\bibnamefont {Gross}},\
  and\ \bibinfo {author} {\bibfnamefont {O.}~\bibnamefont {Lebedev}},\
  }\bibfield  {title} {\bibinfo {title} {{A second Higgs from the Higgs
  portal}},\ }\href {https://doi.org/10.1007/JHEP05(2015)057} {\bibfield
  {journal} {\bibinfo  {journal} {JHEP}\ }\textbf {\bibinfo {volume} {05}},\
  \bibinfo {pages} {057}},\ \Eprint {https://arxiv.org/abs/1502.01361}
  {arXiv:1502.01361 [hep-ph]} \BibitemShut {NoStop}%
\bibitem [{\citenamefont {Lindner}\ \emph {et~al.}(2014)\citenamefont
  {Lindner}, \citenamefont {Schmidt},\ and\ \citenamefont
  {Watanabe}}]{Lindner:2013awa}%
  \BibitemOpen
  \bibfield  {author} {\bibinfo {author} {\bibfnamefont {M.}~\bibnamefont
  {Lindner}}, \bibinfo {author} {\bibfnamefont {D.}~\bibnamefont {Schmidt}},\
  and\ \bibinfo {author} {\bibfnamefont {A.}~\bibnamefont {Watanabe}},\
  }\bibfield  {title} {\bibinfo {title} {{Dark matter and U$(1)'$ symmetry for
  the right-handed neutrinos}},\ }\href
  {https://doi.org/10.1103/PhysRevD.89.013007} {\bibfield  {journal} {\bibinfo
  {journal} {Phys. Rev. D}\ }\textbf {\bibinfo {volume} {89}},\ \bibinfo
  {pages} {013007} (\bibinfo {year} {2014})},\ \Eprint
  {https://arxiv.org/abs/1310.6582} {arXiv:1310.6582 [hep-ph]} \BibitemShut
  {NoStop}%
\bibitem [{ATL()}]{ATLAS:SUSY}%
  \BibitemOpen
  \href@noop {} {}\bibinfo {howpublished}
  {\url{https://atlas.web.cern.ch/Atlas/GROUPS/PHYSICS/PUBNOTES/ATL-PHYS-PUB-2021-019/}}\BibitemShut
  {NoStop}%
\bibitem [{CMS()}]{CMS:SUSY}%
  \BibitemOpen
  \href@noop {} {}\bibinfo {howpublished}
  {\url{http://cms-results.web.cern.ch/cms-results/public-results/publications/SUS/SUS.html}}\BibitemShut
  {NoStop}%
\bibitem [{\citenamefont {Grzadkowski}\ \emph {et~al.}(2010)\citenamefont
  {Grzadkowski}, \citenamefont {Iskrzynski}, \citenamefont {Misiak},\ and\
  \citenamefont {Rosiek}}]{Grzadkowski:2010es}%
  \BibitemOpen
  \bibfield  {author} {\bibinfo {author} {\bibfnamefont {B.}~\bibnamefont
  {Grzadkowski}}, \bibinfo {author} {\bibfnamefont {M.}~\bibnamefont
  {Iskrzynski}}, \bibinfo {author} {\bibfnamefont {M.}~\bibnamefont {Misiak}},\
  and\ \bibinfo {author} {\bibfnamefont {J.}~\bibnamefont {Rosiek}},\
  }\bibfield  {title} {\bibinfo {title} {{Dimension-Six Terms in the Standard
  Model Lagrangian}},\ }\href {https://doi.org/10.1007/JHEP10(2010)085}
  {\bibfield  {journal} {\bibinfo  {journal} {JHEP}\ }\textbf {\bibinfo
  {volume} {10}},\ \bibinfo {pages} {085}},\ \Eprint
  {https://arxiv.org/abs/1008.4884} {arXiv:1008.4884 [hep-ph]} \BibitemShut
  {NoStop}%
\bibitem [{\citenamefont {Tr\'ocs\'anyi}(2020)}]{Trocsanyi:2018bkm}%
  \BibitemOpen
  \bibfield  {author} {\bibinfo {author} {\bibfnamefont {Z.}~\bibnamefont
  {Tr\'ocs\'anyi}},\ }\bibfield  {title} {\bibinfo {title} {{Super-weak force
  and neutrino masses}},\ }\href {https://doi.org/10.3390/sym12010107}
  {\bibfield  {journal} {\bibinfo  {journal} {Symmetry}\ }\textbf {\bibinfo
  {volume} {12}},\ \bibinfo {pages} {107} (\bibinfo {year} {2020})},\ \Eprint
  {https://arxiv.org/abs/1812.11189} {arXiv:1812.11189 [hep-ph]} \BibitemShut
  {NoStop}%
\bibitem [{\citenamefont {Iwamoto}\ \emph {et~al.}(2021)\citenamefont
  {Iwamoto}, \citenamefont {K\"arkk\"ainen}, \citenamefont {P\'eli},\ and\
  \citenamefont {Tr\'ocs\'anyi}}]{Iwamoto:2021wko}%
  \BibitemOpen
  \bibfield  {author} {\bibinfo {author} {\bibfnamefont {S.}~\bibnamefont
  {Iwamoto}}, \bibinfo {author} {\bibfnamefont {T.~J.}\ \bibnamefont
  {K\"arkk\"ainen}}, \bibinfo {author} {\bibfnamefont {Z.}~\bibnamefont
  {P\'eli}},\ and\ \bibinfo {author} {\bibfnamefont {Z.}~\bibnamefont
  {Tr\'ocs\'anyi}},\ }\bibfield  {title} {\bibinfo {title} {{One-loop
  corrections to light neutrino masses in gauged U(1) extensions of the
  standard model}},\ }\href {https://doi.org/10.1103/PhysRevD.104.055042}
  {\bibfield  {journal} {\bibinfo  {journal} {Phys. Rev. D}\ }\textbf {\bibinfo
  {volume} {104}},\ \bibinfo {pages} {055042} (\bibinfo {year} {2021})},\
  \Eprint {https://arxiv.org/abs/2104.14571} {arXiv:2104.14571 [hep-ph]}
  \BibitemShut {NoStop}%
\bibitem [{\citenamefont {Iwamoto}\ \emph {et~al.}(2022)\citenamefont
  {Iwamoto}, \citenamefont {Seller},\ and\ \citenamefont
  {Tr\'ocs\'anyi}}]{Iwamoto:2021fup}%
  \BibitemOpen
  \bibfield  {author} {\bibinfo {author} {\bibfnamefont {S.}~\bibnamefont
  {Iwamoto}}, \bibinfo {author} {\bibfnamefont {K.}~\bibnamefont {Seller}},\
  and\ \bibinfo {author} {\bibfnamefont {Z.}~\bibnamefont {Tr\'ocs\'anyi}},\
  }\bibfield  {title} {\bibinfo {title} {{Sterile neutrino dark matter in a
  U(1) extension of the standard model}},\ }\href
  {https://doi.org/10.1088/1475-7516/2022/01/035} {\bibfield  {journal}
  {\bibinfo  {journal} {JCAP}\ }\textbf {\bibinfo {volume} {01}}\bibfield
  {number} {\bibinfo  {number} { (01)},\ \bibinfo {pages} {035}},\ }\Eprint
  {https://arxiv.org/abs/2104.11248} {arXiv:2104.11248 [hep-ph]} \BibitemShut
  {NoStop}%
\bibitem [{\citenamefont {P\'eli}\ \emph {et~al.}(2020)\citenamefont {P\'eli},
  \citenamefont {N\'andori},\ and\ \citenamefont
  {Tr\'ocs\'anyi}}]{Peli:2019vtp}%
  \BibitemOpen
  \bibfield  {author} {\bibinfo {author} {\bibfnamefont {Z.}~\bibnamefont
  {P\'eli}}, \bibinfo {author} {\bibfnamefont {I.}~\bibnamefont {N\'andori}},\
  and\ \bibinfo {author} {\bibfnamefont {Z.}~\bibnamefont {Tr\'ocs\'anyi}},\
  }\bibfield  {title} {\bibinfo {title} {{Particle physics model of curvaton
  inflation in a stable universe}},\ }\href
  {https://doi.org/10.1103/PhysRevD.101.063533} {\bibfield  {journal} {\bibinfo
   {journal} {Phys. Rev. D}\ }\textbf {\bibinfo {volume} {101}},\ \bibinfo
  {pages} {063533} (\bibinfo {year} {2020})},\ \Eprint
  {https://arxiv.org/abs/1911.07082} {arXiv:1911.07082 [hep-ph]} \BibitemShut
  {NoStop}%
\bibitem [{Note1()}]{Note1}%
  \BibitemOpen
  \bibinfo {note} {We shall denote the pole mass of a particle $p$ as
  $M_p$.}\BibitemShut {Stop}%
\bibitem [{\citenamefont {Zyla}\ \emph {et~al.}(2020)\citenamefont {Zyla} \emph
  {et~al.}}]{ParticleDataGroup:2020ssz}%
  \BibitemOpen
  \bibfield  {author} {\bibinfo {author} {\bibfnamefont {P.~A.}\ \bibnamefont
  {Zyla}} \emph {et~al.} (\bibinfo {collaboration} {Particle Data Group}),\
  }\bibfield  {title} {\bibinfo {title} {{Review of Particle Physics}},\ }\href
  {https://doi.org/10.1093/ptep/ptaa104} {\bibfield  {journal} {\bibinfo
  {journal} {PTEP}\ }\textbf {\bibinfo {volume} {2020}},\ \bibinfo {pages}
  {083C01} (\bibinfo {year} {2020})}\BibitemShut {NoStop}%
\bibitem [{\citenamefont {Banerjee}\ \emph {et~al.}(2019)\citenamefont
  {Banerjee} \emph {et~al.}}]{NA64:2019imj}%
  \BibitemOpen
  \bibfield  {author} {\bibinfo {author} {\bibfnamefont {D.}~\bibnamefont
  {Banerjee}} \emph {et~al.} (\bibinfo {collaboration} {{NA64}}),\ }\bibfield
  {title} {\bibinfo {title} {{Dark matter search in missing energy events with
  NA64}},\ }\href {https://doi.org/10.1103/PhysRevLett.123.121801} {\bibfield
  {journal} {\bibinfo  {journal} {Phys. Rev. Lett.}\ }\textbf {\bibinfo
  {volume} {123}},\ \bibinfo {pages} {121801} (\bibinfo {year} {2019})},\
  \Eprint {https://arxiv.org/abs/1906.00176} {arXiv:1906.00176 [hep-ex]}
  \BibitemShut {NoStop}%
\bibitem [{\citenamefont {Porod}(2003)}]{Porod:2003um}%
  \BibitemOpen
  \bibfield  {author} {\bibinfo {author} {\bibfnamefont {W.}~\bibnamefont
  {Porod}},\ }\bibfield  {title} {\bibinfo {title} {{SPheno, a program for
  calculating supersymmetric spectra, SUSY particle decays and SUSY particle
  production at e+ e- colliders}},\ }\href
  {https://doi.org/10.1016/S0010-4655(03)00222-4} {\bibfield  {journal}
  {\bibinfo  {journal} {Comput. Phys. Commun.}\ }\textbf {\bibinfo {volume}
  {153}},\ \bibinfo {pages} {275} (\bibinfo {year} {2003})},\ \Eprint
  {https://arxiv.org/abs/hep-ph/0301101} {arXiv:hep-ph/0301101} \BibitemShut
  {NoStop}%
\bibitem [{\citenamefont {Porod}\ and\ \citenamefont
  {Staub}(2012)}]{Porod:2011nf}%
  \BibitemOpen
  \bibfield  {author} {\bibinfo {author} {\bibfnamefont {W.}~\bibnamefont
  {Porod}}\ and\ \bibinfo {author} {\bibfnamefont {F.}~\bibnamefont {Staub}},\
  }\bibfield  {title} {\bibinfo {title} {{SPheno 3.1: Extensions including
  flavour, CP-phases and models beyond the MSSM}},\ }\href
  {https://doi.org/10.1016/j.cpc.2012.05.021} {\bibfield  {journal} {\bibinfo
  {journal} {Comput. Phys. Commun.}\ }\textbf {\bibinfo {volume} {183}},\
  \bibinfo {pages} {2458} (\bibinfo {year} {2012})},\ \Eprint
  {https://arxiv.org/abs/1104.1573} {arXiv:1104.1573 [hep-ph]} \BibitemShut
  {NoStop}%
\bibitem [{Note2()}]{Note2}%
  \BibitemOpen
  \bibinfo {note} {There is a price to pay for this speed-up, namely we discard
  small portions of the parameter space, where $w^{(1)}(\protect \ensuremath
  {{M}_{\protect \rm t}})$ is not positive but $w^{(2)}(\protect \ensuremath
  {{M}_{\protect \rm t}})$ is so.}\BibitemShut {Stop}%
\bibitem [{\citenamefont {Elias-Miró}\ \emph {et~al.}(2012)\citenamefont
  {Elias-Miró}, \citenamefont {Espinosa}, \citenamefont {Giudice},
  \citenamefont {Lee},\ and\ \citenamefont {Strumia}}]{SM-eft-stability}%
  \BibitemOpen
  \bibfield  {author} {\bibinfo {author} {\bibfnamefont {J.}~\bibnamefont
  {Elias-Miró}}, \bibinfo {author} {\bibfnamefont {J.~R.}\ \bibnamefont
  {Espinosa}}, \bibinfo {author} {\bibfnamefont {G.~F.}\ \bibnamefont
  {Giudice}}, \bibinfo {author} {\bibfnamefont {H.~M.}\ \bibnamefont {Lee}},\
  and\ \bibinfo {author} {\bibfnamefont {A.}~\bibnamefont {Strumia}},\
  }\bibfield  {title} {\bibinfo {title} {{Stabilization of the electroweak
  vacuum by a scalar threshold effect}},\ }\href
  {https://doi.org/10.1007/JHEP06(2012)031} {\bibfield  {journal} {\bibinfo
  {journal} {JHEP}\ }\textbf {\bibinfo {volume} {2012}}\bibfield  {number}
  {\bibinfo  {number} { (31)},\ \bibinfo {pages} {18}},\ }\Eprint
  {https://arxiv.org/abs/1203.0237} {1203.0237} \BibitemShut {NoStop}%
\bibitem [{\citenamefont {Robens}\ and\ \citenamefont
  {Stefaniak}(2015)}]{Robens:2015gla}%
  \BibitemOpen
  \bibfield  {author} {\bibinfo {author} {\bibfnamefont {T.}~\bibnamefont
  {Robens}}\ and\ \bibinfo {author} {\bibfnamefont {T.}~\bibnamefont
  {Stefaniak}},\ }\bibfield  {title} {\bibinfo {title} {{Status of the Higgs
  Singlet Extension of the Standard Model after LHC Run 1}},\ }\href
  {https://doi.org/10.1140/epjc/s10052-015-3323-y} {\bibfield  {journal}
  {\bibinfo  {journal} {Eur. Phys. J. C}\ }\textbf {\bibinfo {volume} {75}},\
  \bibinfo {pages} {104} (\bibinfo {year} {2015})},\ \Eprint
  {https://arxiv.org/abs/1501.02234} {arXiv:1501.02234 [hep-ph]} \BibitemShut
  {NoStop}%
\bibitem [{\citenamefont {Baak}\ \emph {et~al.}(2012)\citenamefont {Baak},
  \citenamefont {Goebel}, \citenamefont {Haller}, \citenamefont {Hoecker},
  \citenamefont {Kennedy}, \citenamefont {Kogler}, \citenamefont {Moenig},
  \citenamefont {Schott},\ and\ \citenamefont {Stelzer}}]{Baak:2012kk}%
  \BibitemOpen
  \bibfield  {author} {\bibinfo {author} {\bibfnamefont {M.}~\bibnamefont
  {Baak}}, \bibinfo {author} {\bibfnamefont {M.}~\bibnamefont {Goebel}},
  \bibinfo {author} {\bibfnamefont {J.}~\bibnamefont {Haller}}, \bibinfo
  {author} {\bibfnamefont {A.}~\bibnamefont {Hoecker}}, \bibinfo {author}
  {\bibfnamefont {D.}~\bibnamefont {Kennedy}}, \bibinfo {author} {\bibfnamefont
  {R.}~\bibnamefont {Kogler}}, \bibinfo {author} {\bibfnamefont
  {K.}~\bibnamefont {Moenig}}, \bibinfo {author} {\bibfnamefont
  {M.}~\bibnamefont {Schott}},\ and\ \bibinfo {author} {\bibfnamefont
  {J.}~\bibnamefont {Stelzer}},\ }\bibfield  {title} {\bibinfo {title} {{The
  Electroweak Fit of the Standard Model after the Discovery of a New Boson at
  the LHC}},\ }\href {https://doi.org/10.1140/epjc/s10052-012-2205-9}
  {\bibfield  {journal} {\bibinfo  {journal} {Eur. Phys. J. C}\ }\textbf
  {\bibinfo {volume} {72}},\ \bibinfo {pages} {2205} (\bibinfo {year}
  {2012})},\ \Eprint {https://arxiv.org/abs/1209.2716} {arXiv:1209.2716
  [hep-ph]} \BibitemShut {NoStop}%
\bibitem [{\citenamefont {Ciuchini}\ \emph {et~al.}(2013)\citenamefont
  {Ciuchini}, \citenamefont {Franco}, \citenamefont {Mishima},\ and\
  \citenamefont {Silvestrini}}]{Ciuchini:2013pca}%
  \BibitemOpen
  \bibfield  {author} {\bibinfo {author} {\bibfnamefont {M.}~\bibnamefont
  {Ciuchini}}, \bibinfo {author} {\bibfnamefont {E.}~\bibnamefont {Franco}},
  \bibinfo {author} {\bibfnamefont {S.}~\bibnamefont {Mishima}},\ and\ \bibinfo
  {author} {\bibfnamefont {L.}~\bibnamefont {Silvestrini}},\ }\bibfield
  {title} {\bibinfo {title} {{Electroweak Precision Observables, New Physics
  and the Nature of a 126 GeV Higgs Boson}},\ }\href
  {https://doi.org/10.1007/JHEP08(2013)106} {\bibfield  {journal} {\bibinfo
  {journal} {JHEP}\ }\textbf {\bibinfo {volume} {08}},\ \bibinfo {pages}
  {106}},\ \Eprint {https://arxiv.org/abs/1306.4644} {arXiv:1306.4644 [hep-ph]}
  \BibitemShut {NoStop}%
\bibitem [{\citenamefont {Degrassi}\ \emph {et~al.}(2014)\citenamefont
  {Degrassi}, \citenamefont {Gambino},\ and\ \citenamefont
  {Giardino}}]{degrassi2014}%
  \BibitemOpen
  \bibfield  {author} {\bibinfo {author} {\bibfnamefont {G.}~\bibnamefont
  {Degrassi}}, \bibinfo {author} {\bibfnamefont {P.}~\bibnamefont {Gambino}},\
  and\ \bibinfo {author} {\bibfnamefont {P.~P.}\ \bibnamefont {Giardino}},\
  }\href@noop {} {\bibinfo {title} {The $m_{\scriptscriptstyle
  w}-m_{\scriptscriptstyle z}$ interdependence in the standard model: a new
  scrutiny}} (\bibinfo {year} {2014}),\ \Eprint
  {https://arxiv.org/abs/1411.7040} {arXiv:1411.7040 [hep-ph]} \BibitemShut
  {NoStop}%
\bibitem [{\citenamefont {Fernandez-Martinez}\ \emph
  {et~al.}(2016)\citenamefont {Fernandez-Martinez}, \citenamefont
  {Hernandez-Garcia},\ and\ \citenamefont
  {Lopez-Pavon}}]{Fernandez-Martinez:2016lgt}%
  \BibitemOpen
  \bibfield  {author} {\bibinfo {author} {\bibfnamefont {E.}~\bibnamefont
  {Fernandez-Martinez}}, \bibinfo {author} {\bibfnamefont {J.}~\bibnamefont
  {Hernandez-Garcia}},\ and\ \bibinfo {author} {\bibfnamefont {J.}~\bibnamefont
  {Lopez-Pavon}},\ }\bibfield  {title} {\bibinfo {title} {{Global constraints
  on heavy neutrino mixing}},\ }\href {https://doi.org/10.1007/JHEP08(2016)033}
  {\bibfield  {journal} {\bibinfo  {journal} {JHEP}\ }\textbf {\bibinfo
  {volume} {08}},\ \bibinfo {pages} {033}},\ \Eprint
  {https://arxiv.org/abs/1605.08774} {arXiv:1605.08774 [hep-ph]} \BibitemShut
  {NoStop}%
\bibitem [{\citenamefont {Blennow}\ \emph {et~al.}(2022)\citenamefont
  {Blennow}, \citenamefont {Coloma}, \citenamefont
  {Fern\'andez-Mart\'\i{}nez},\ and\ \citenamefont
  {Gonz\'alez-L\'opez}}]{Blennow:2022yfm}%
  \BibitemOpen
  \bibfield  {author} {\bibinfo {author} {\bibfnamefont {M.}~\bibnamefont
  {Blennow}}, \bibinfo {author} {\bibfnamefont {P.}~\bibnamefont {Coloma}},
  \bibinfo {author} {\bibfnamefont {E.}~\bibnamefont
  {Fern\'andez-Mart\'\i{}nez}},\ and\ \bibinfo {author} {\bibfnamefont
  {M.}~\bibnamefont {Gonz\'alez-L\'opez}},\ }\bibfield  {title} {\bibinfo
  {title} {{Right-handed neutrinos and the CDF II anomaly}},\ }\href@noop {} {\
   (\bibinfo {year} {2022})},\ \Eprint {https://arxiv.org/abs/2204.04559}
  {arXiv:2204.04559 [hep-ph]} \BibitemShut {NoStop}%
\bibitem [{\citenamefont {L\'opez-Val}\ and\ \citenamefont
  {Robens}(2014)}]{Lopez-Val:2014jva}%
  \BibitemOpen
  \bibfield  {author} {\bibinfo {author} {\bibfnamefont {D.}~\bibnamefont
  {L\'opez-Val}}\ and\ \bibinfo {author} {\bibfnamefont {T.}~\bibnamefont
  {Robens}},\ }\bibfield  {title} {\bibinfo {title} {{\ensuremath{\Delta}r and
  the W-boson mass in the singlet extension of the standard model}},\ }\href
  {https://doi.org/10.1103/PhysRevD.90.114018} {\bibfield  {journal} {\bibinfo
  {journal} {Phys. Rev. D}\ }\textbf {\bibinfo {volume} {90}},\ \bibinfo
  {pages} {114018} (\bibinfo {year} {2014})},\ \Eprint
  {https://arxiv.org/abs/1406.1043} {arXiv:1406.1043 [hep-ph]} \BibitemShut
  {NoStop}%
\bibitem [{\citenamefont {Aaltonen}\ \emph {et~al.}(2022)\citenamefont
  {Aaltonen} \emph {et~al.}}]{CDF:2022hxs}%
  \BibitemOpen
  \bibfield  {author} {\bibinfo {author} {\bibfnamefont {T.}~\bibnamefont
  {Aaltonen}} \emph {et~al.} (\bibinfo {collaboration} {CDF}),\ }\bibfield
  {title} {\bibinfo {title} {{High-precision measurement of the $W$ boson mass
  with the CDF II detector}},\ }\href {https://doi.org/10.1126/science.abk1781}
  {\bibfield  {journal} {\bibinfo  {journal} {Science}\ }\textbf {\bibinfo
  {volume} {376}},\ \bibinfo {pages} {170} (\bibinfo {year}
  {2022})}\BibitemShut {NoStop}%
\end{thebibliography}

\providecommand{\noopsort}[1]{}\providecommand{\singleletter}[1]{#1}%

\end{document}